\documentclass{aa}  

\usepackage{graphicx}
\usepackage{txfonts}
\usepackage{color}
\usepackage{xcolor}
\definecolor{xlinkcolor}{cmyk}{1,1,0,0}
\usepackage[bookmarks=true,         
bookmarksopen=true,
pdfnewwindow=true,      
colorlinks=true,    
linkcolor=xlinkcolor,     
citecolor=xlinkcolor,     
filecolor=xlinkcolor,  
urlcolor=xlinkcolor,      
final=true]{hyperref} 
\hypersetup{pdfstartview=Fit}
\usepackage{bookmark}
\usepackage[normalem]{ulem}
\makeatletter
\renewcommand*\aa@pageof{, page \thepage{} of \pageref*{LastPage}}
\makeatother

\begin{document}

   \title{The robustness of inferred envelope and core rotation rates of red-giant stars from asteroseismology }

   \author{F. Ahlborn \inst{1,2}
          \and E. P. Bellinger \inst{3,2,4}
          \and S. Hekker \inst{1,5,4}
          \and S. Basu \inst{3}
          \and D. Mokrytska \inst{1,5}}
   \institute{Heidelberger Institut f\"ur Theoretische Studien, Schloss-Wolfsbrunnenweg 35, 69118 Heidelberg, Germany\\
              \email{felix.ahlborn@h-its.org}
    \and
        Max-Planck-Institut f\"ur Astrophysik, Karl-Schwarzschild-Stra{\ss}e 1, 85748 Garching, Germany
    \and 
        Department of Astronomy, Yale University, New Haven, CT 06520, USA
    \and         
        Stellar Astrophysics Centre, Department of Physics and Astronomy, Aarhus University, Ny Munkegade 120, DK-8000 Aarhus C, Denmark
    \and
        Center for Astronomy (ZAH/LSW), Heidelberg University, Königstuhl 12, 69117 Heidelberg, Germany
             }

   \date{Received; accepted}

 
  \abstract
   {Rotation is an important, yet poorly-modelled phenomenon of stellar structure and evolution. Accurate estimates of internal rotation rates are therefore valuable for constraining stellar evolution models.} 
   {We aim to assess the accuracy of asteroseismic estimates of internal rotation rates and how these depend on the fundamental stellar parameters.}
   {We apply the recently-developed method called extended-MOLA inversions to infer localised estimates of internal rotation rates of synthetic observations of red giants. We search for suitable reference stellar models following a grid-based approach, and assess the robustness of the resulting inferences to the choice of reference model.}
   {We find that matching the mixed mode pattern between the observation and the reference model is an important criterion to select suitable reference models. We propose to i) select a set of reference models based on the correlation between the observed rotational splittings and the mode-trapping parameter ii) compute rotation rates for all these models iii) use the mean value obtained across the whole set as the estimate of the internal rotation rates. We find that the effect of a near surface perturbation in the synthetic observations on the rotation rates estimated based on the correlation between the observed rotational splittings and the mode-trapping parameter is negligible.}
   {We conclude that when using an ensemble of reference models, constructed based on matching the mixed mode pattern, the input rotation rates can be recovered across a range of fundamental stellar parameters like mass, mixing-length parameter and composition. Further, red-giant rotation rates determined in this way are also independent of a near surface perturbation of stellar structure.}

   \keywords{asteroseismology -- stars: rotation -- stars: oscillations -- stars: interiors}

   \maketitle
%

\section{Introduction}
\label{secintro}
    
    Stars form from gas clouds that already have some angular momentum. As a consequence, every star is expected to rotate. This in turn gives rise to several physical processes, such as rotationally-induced mixing of chemical elements, meridional circulation, and various other instabilities \citep{maeder2009}. These processes can significantly change the overall stellar structure and thus the evolution of the stars \citep{eggenberger2010}, for example, prolonging their lifetimes. However, including the effects and the evolution of the stellar rotation in stellar structure and evolution modelling presents both numerical and theoretical challenges. Especially in the subgiant and red-giant evolutionary phases, current theoretical models of rotation cannot reproduce observed stellar rotation rates \citep{aerts2019}. Due to a lack of efficient angular-momentum transport mechanisms, theoretical stellar models predict ratios of core to surface rotation rates for subgiant and red-giant stars that are orders of magnitude higher than observed \citep{eggenberger2012, marques2013, ceillier2013, cantiello2014}. In order to resolve this discrepancy, other means of angular momentum transport have been considered, such as mixed modes \citep{belkacem2015a, belkacem2015b}, internal gravity waves \citep{alvan2013, fuller2014,pincon2017}, and magnetism \citep{spruit2002,cantiello2014,fuller2019, eggenberger2019}. As asteroseismic  measurements of envelope rotation rates have been made only for a limited set of stars up to now, additional observations are necessary to discriminate between the different scenarios.
    
    In red-giant stars global oscillations are stochastically excited by turbulent convection \citep{kjeldsen1995, bouchy2001}. These oscillation modes typically propagate in two cavities, which are the {\it g}-mode cavity in which the restoring force is buoyancy, and the {\it p}-mode cavity in which the restoring force is the pressure gradient. Different oscillation modes probe different depths of the star, and thus from the observation of many modes, one can draw conclusions about the internal stellar structure. In the case of red giants, however, all non-radial modes have a mixed nature, that is, they behave as gravity modes in deep interiors, and as acoustic modes in the envelope. This coupling of the {\it g}- and {\it p}-modes allows us to probe the conditions in red-giant cores \citep[e.g.][]{dupret2009, beck2011, bedding2011}.
          \begin{figure*}
   \centering
   \includegraphics{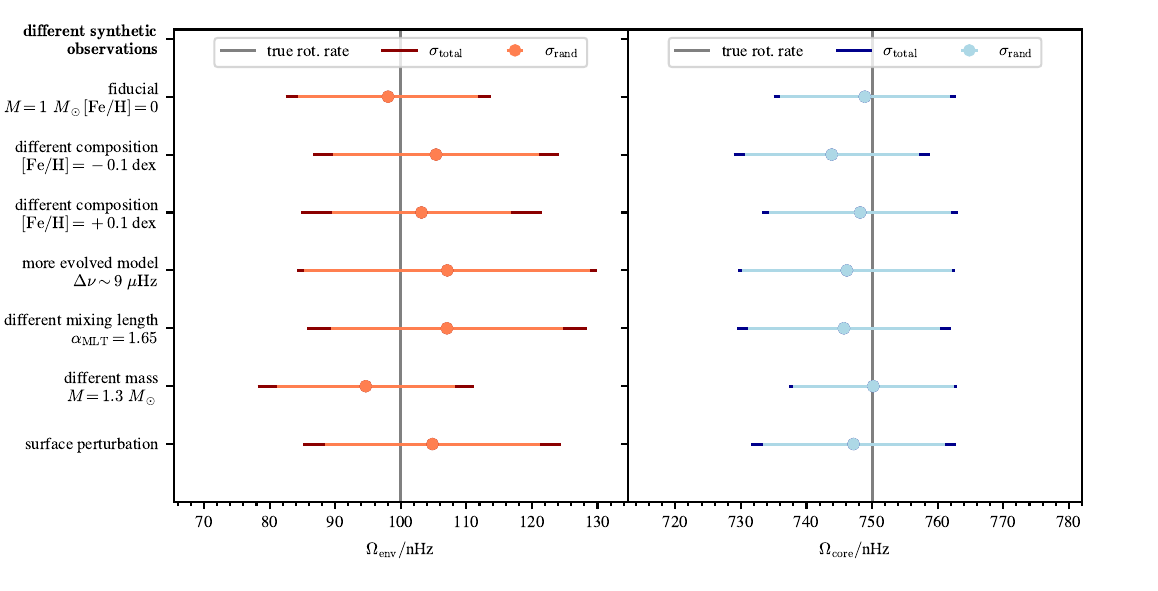}
      \caption{Envelope (left) and core (right) rotation rates estimated using an ensemble of reference models for different synthetic observations of red-giant stars. The comparison with the input rotation rates shows that we are able to recover the underlying rotation rates in all of the synthetic observations. As a fiducial synthetic observation we used a 1~M$_\odot$ model with solar metallicity, a mixing length parameter of 1.8 and a large frequency separation of $\Delta\nu\sim15~\mu$Hz. For the other synthetic observations we varied the parameters indicated by the label (see also Table~\ref{tabsynobs}). The numerical values of rotation rates and uncertainties are summarised in Table~\ref{tabresults}. The error given in the dark colours in each panel is calculated from the random and reference model uncertainties by error propagation. The error bar in the light colours is representing the contribution of the random error $\sigma_{\rm rand}$ alone. The vertical grey lines indicate the input values.}
         \label{figsummary}
   \end{figure*}
    To measure the internal rotation rates by means of asteroseismology, we make use of the effect of rotation on the eigenmodes of the oscillations. In non-rotating stars, oscillation modes are degenerate in their azimuthal order. Rotation lifts this degeneracy and splits an oscillation mode into standing and travelling pro- and retrograde components. In the power spectrum, this becomes visible as the so-called `rotational splitting' of oscillation mode frequencies. To exploit the full potential of the observed splittings and infer the internal red-giant rotation rates, theoretical sensitivity functions of the splittings need to be obtained from reference stellar models. However, some limited amount of information on the internal rotation can be deduced solely from observations without the need for a reference stellar model. The analysis of the {\it g}-dominated modes allows getting a model independent estimate of the mean core rotation rates of red giants \citep{beck2012,mosser2012,gehan2018, mosser2024}. To take into account the varying ratio of core and envelope sensitivity in mixed modes, \cite{goupil2013} described the rotational splittings as a linear function of the mode trapping parameter $\zeta$--- the ratio of the core inertia to the total inertia ---enabling the determination of average core and envelope rotation rates. This got further developed by \cite{deheuvels2015}, who provided an estimate for the mode trapping parameter that allows for a model-independent estimate of the internal rotation rates. 
    
    As an alternative to using the linear relation between the mode trapping parameter and the rotational splittings, we use so-called rotational inversions, relying on a linear perturbative expansion of the rotationally split mode frequencies. To compute a rotational inversion, it is necessary to construct a reference stellar model that reproduces the internal structure of the observed star as close as possible, including mode frequencies and non-seismic observables. Here, special care has to be taken with the variation of the {\it p}- and {\it g}-mode nature of the modes. Such an approach has been adopted in studies of sub-giants \citep{deheuvels2012, deheuvels2014, deheuvels2020, buldgen2024}, red-giant branch stars \citep{dimauro2016, dimauro2018, triana2017, beck2014, beck2018, fellay2021} and secondary red clump stars \citep{deheuvels2015}. Numerous studies have compared the results of different methods for determining the internal rotation rates and have shown that these methods are in agreement with each other \citep{christensen1990, schou1998, deheuvels2012, dimauro2016}.
    
    Applying the original multiplicative optimally localised averages (MOLA) inversions \citep{backus1968} to determine internal rotation rates, \cite{ahlborn2020} showed that estimated envelope rotation rates of red-giant stars suffer from substantial relative systematic errors that can be up to about 200\% for stars close to the luminosity bump on the red-giant branch (RGB). Recently, \cite{ahlborn2022} proposed an extension to the original MOLA method, called extended MOLA (eMOLA), which enables the construction of surface averaging kernels with virtually no cumulative sensitivity to the rotation rate of the core. Using eMOLA inversions, the relative systematic errors due to the inversion method can be essentially suppressed. As described above, rotational inversions rely on the theoretical sensitivity functions obtained from stellar reference models. Therefore, discrepancies between the reference model and the observed star constitute another source of uncertainty for the estimated rotation rates. Previous studies showed that similar rotation rates were estimated with reference models of different mass \citep{deheuvels2012, dimauro2016}. \cite{dimauro2016} invert for the same set of observed rotational splittings with two reference models of masses of $M_1=1.02~M_\odot$ and $M_2=1.13~M_\odot$, and find that the estimated rotation rates agree within uncertainties despite the difference in mass.
    
    Finally, deviations in the mode frequencies due to inadequate modelling of the near surface layers in the reference models, also known as the `surface effect' \citep{brown1984, christensen1988}, may impact the determination of the internal rotation rates \citep{ong2024}. Different ways have been proposed to mitigate this surface effect, consisting of parameterisations of the observed frequency deviations \citep{kjeldsen2008, ball2014}, improving the modelling of the near surface structure of the models \citep[][and references therein]{jorgensen2018} or modifying the computation of the oscillation mode frequencies \citep{houdek2017}. Mitigating the surface effect becomes especially difficult for mixed modes, as the difference in the near surface layers only acts on the {\it p}-mode component of the mixed mode, while the {\it g}-mode component remains unaffected \citep{ball2018, ong2021}. Due to the perturbation of the {\it p}-mode frequency caused by the perturbation of the near surface layers, a different {\it g}-mode couples with this {\it p}-mode as compared to the case without a near surface perturbation. This leads to the formation of a different mixed mode, with a different sensitivity to the internal rotation \citep{ong2024}.

    In this study, we investigate the sensitivity of inferred rotation profiles to the choice of the reference model, with the goal to understand how good a stellar reference model needs to be in order to give reliable information on the internal rotation profile in a red-giant star and how these internal rotation rates depend on the properties of the observed star. For comparison with previous studies, we used an OLA method (eMOLA) to compute the rotational inversions and largely assumed a two-zonal configuration of the internal rotation profile. We visualise the main result of this work in Fig.~\ref{figsummary} where we show estimated envelope and core rotation rates for different synthetic observations. The comparison with the input rotation rates shows that we are able to recover the underlying rotation rates in all of the synthetic observations. This shows that we are able to obtain an unbiased estimate of the internal rotation rates largely independent of the fundamental stellar parameters. We find that the uncertainties introduced due to a discrepant structure between the observed star and the reference models are of similar order as the uncertainties introduced by measurement errors. This allows us to obtain well constrained envelope rotation rates and will be an important probe for theories of angular momentum transport. The following text is dedicated to discussing how we arrived at the results shown in Fig.~\ref{figsummary}.

    \section{Methods and synthetic data}\label{secmethods}
    To estimate the internal rotation rates of red-giant stars we use rotational inversions, which we describe in Sect.~\ref{secrotinv}. The reference models needed to compute these rotational inversions are selected from the grids of stellar models described in Sect.~\ref{secstellarmodels}. We assess the accuracy of the rotational inversion results by constructing different sets of synthetic observations with known input parameters, as described in Sect.~\ref{secsynth}. To test the impact of the surface effect on the rotational inversions, we construct a set of synthetic observations from a stellar model that includes a surface perturbation as described in Sect.~\ref{secsurfpert}.
    
    \subsection{Rotational inversions}\label{secrotinv}
    Commonly-used rotational inversion methods include the methods of optimally localised averages (OLA) and regularised least squares (RLS). The class of OLA methods relies on the construction of localised functions, called averaging kernels, that are used to compute an estimate of the rotation rate at a so-called target radius. So far, mainly the multiplicative \citep[MOLA,][]{backus1968} and subtractive \citep[SOLA,][]{pijpers1992, pijpers1994} OLA have been used to estimate internal rotation rates of stars \citep{christensen1990, schou1998, deheuvels2012, deheuvels2014, dimauro2016, triana2017}. \cite{ahlborn2020} showed that envelope rotation rates obtained from MOLA inversions suffer from substantial systematic errors, especially for more evolved red giants. \cite{ahlborn2022} therefore proposed a new rotational inversion method, called extended MOLA (eMOLA), that eliminates these systematic errors.

    To obtain the averaging kernels, the eMOLA inversion method uses the following objective function: 

    \begin{align}
        Z_\text{eMOLA}&=\int_0^RK(r,r_0)^2J(r,r_0)\,\text{d}r\nonumber\\
        &+\theta\left[\int_0^RK(r,r_0)J(r,r_0)\,\text{d}r\right]^2\nonumber\\
        &+\frac{\mu}{\mu_0}\sigma^2_{\Omega(r_0)}.
    \end{align}
    where $K(r, r_0)$ refers to the averaging kernel localised at the target radius $r_0$, the function $J(r, r_0)$ is weighting the sensitivity of the averaging kernel and $\theta$ is a parameter balancing the first and the second term. The first term, which minimises the amplitude of the averaging kernel away from the target radius, is also part of the original MOLA objective function while the second term, which minimises the cumulative sensitivity away from the target radius, was introduced in the eMOLA inversion method. The last term is an error suppression term. We denote the so-called trade-off parameter with $\mu$. This parameter balances the uncertainty of the solution derived from propagated errors $\sigma_{\Omega(r_0)}$ with the resolution of the inversions, as indicated by the width of the averaging kernels. The symbol $\mu_0$ denotes a normalisation constant for the propagated uncertainties. For the details of the rotational inversion methods, we refer the reader to Appendix~\ref{secrotinvapp}.
\subsection{Grids of stellar models}\label{secstellarmodels}
To search for reference models, we construct two different stellar model grids. We construct stellar evolutionary models using Modules for Experiments in Stellar Astrophysics \citep[MESA, version 12778,][]{paxton2011,paxton2013,paxton2015, paxton2018,paxton2019,jermyn2023}. To select the range of stellar parameters to consider in our study, we used the parameters of stars that have been analysed in terms of rotational inversions in \cite{deheuvels2012,dimauro2016} and \cite{triana2017}. The analysis of 16 red-giant stars presented in these studies suggests considering stellar masses of 0.8 up to 2~M$_\odot$. In the first grid we vary the stellar mass in the aforementioned mass range and keep the mixing-length parameter $\alpha_{\rm MLT}$ of convection fixed to a value of 1.8. We refer to this grid as the $M$-grid. In the second grid we also vary the mixing length parameter between 1.5 and 2. We refer to this grid as the $M,\alpha_{\rm MLT}$-grid. For the details of the stellar model grids we refer to Appendix~\ref{secstellarmodelsapp}.
\subsection{Synthetic data}
\label{secsynth}
    To study the extent to which the estimate of the internal rotation profile derived through rotational inversions depends on the choice of a reference model, we start by computing sets of synthetic observations. In this case, we know the underlying rotation profile that we aim to reconstruct using different reference models. To generate the synthetic data, we construct stellar evolutionary models using MESA. For a selected model, we create a set of synthetic observations. Each set consists of global stellar parameters ($\Delta\nu, \nu_{\rm max}$) as well as radial and dipole mode frequencies and rotational kernels and splittings. The global seismic variables are computed from scaling relations \citep{kjeldsen1995} using the reference values from \cite{themessl2018} while the frequencies and kernels are computed using the GYRE oscillation code \citep{townsend2013,townsend2018}. The synthetic observations are supplemented with realistic uncertainties. The details of the synthetic data are described in Appendix~\ref{secsynthapp}. As a synthetic rotation profile we use a step profile with a constant rotation above and below the base of the convection zone (see Fig.~\ref{figrotprofiles}). As core and envelope rotation rates we use $\Omega_{\rm core}=750$~nHz and $\Omega_{\rm core}=100$~nHz, respectively. We refer to this profile as the `envelope step' profile.
    
    We summarise the fundamental parameters of the stellar models used to create the synthetic observations in Table~\ref{tabsynobs}. For the sake of convenience, we use the same microphysics for the stellar models used to create the synthetic data as for the two grids described in Sect.~\ref{secstellarmodels}. When constructing synthetic observations with varying initial metallicity, the initial helium abundance $Y_i$ is determined according to the enrichment law:
    \begin{equation}
        Y_{\rm i} = 0.249 + 1.5 \cdot Z_{\rm i} 
        \label{eqenrich_law}
    \end{equation}
    where we take $Y_{\rm primordial}=0.249$ from \cite{planck} and the ${\rm d}Y/{\rm d}Z=1.5$ from \cite{choi2016} computed from the \cite{asplund2009} protosolar $Y$ and $Z$ values.

\subsection{Surface perturbed models}\label{secsurfpert}
\begin{figure}
    \centering
    \includegraphics{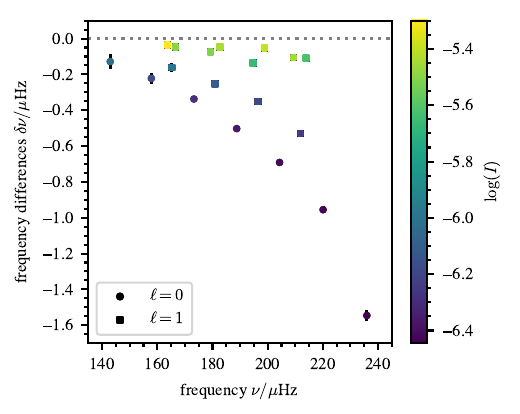}
    \caption{Frequency differences between the surface perturbed model and the fiducial model ($\delta\nu=\nu_{\rm pert}-\nu_{\rm fid}$) as a function of the unperturbed mode frequencies. The mode inertia are colour-coded on a logarithmic scale.}
    \label{figsurfpert}
\end{figure}

To mimic a surface effect, we follow the procedure of \cite{ong2021}. They introduced a perturbation, localised at the surface of the model, to the pressure $p$ and the first adiabatic exponent $\Gamma_1$. We used the original parameters as proposed in \cite{ong2021}. We refer to the model including the perturbation as the `surface perturbed model'. The other fundamental parameters are the same as for the fiducial model (first row of Table~\ref{tabsynobs}). 
The resulting frequency differences between the surface perturbed and the fiducial model for the radial and dipole modes are shown in Fig.~\ref{figsurfpert} as a function of the unperturbed frequencies. Modes that are more sensitive to the surface layers (indicated by lower values of the mode inertia) have a larger frequency difference compared to the unperturbed case. Hence, the radial modes are most affected by the surface perturbation, followed by the {\it p}-dominated dipole modes. The more {\it g}-dominated modes remain mostly unaffected by the surface perturbation. We computed synthetic observations for the surface perturbed model in the same way as described in Sect.~\ref{secsynth}.
\begin{table*}
                \caption{List of stellar models used to generate synthetic observations}
                \label{tabsynobs}
                \centering
                \begin{tabular}{c c c c c c c c c}
                        \hline\hline
                        \rule{0pt}{12pt}name&$M/M_\odot$& $X$&$Y$&$Z$&$\alpha_{\rm MLT}$&$\Delta\nu$&$\nu_{\rm max}$&grid\\
                        \hline
                        \rule{0pt}{10pt}fiducial model &1&0.7155 &0.2703&0.0142&1.8&15.23&197&$M, (M, \alpha_{\rm MLT})$\\
                        \rule{0pt}{10pt} $\left[\frac{{\rm Fe}}{{\rm H}}\right]=-0.1$ dex model &1&0.7226 &0.2660&0.0114&1.8&15.22&195&$(M, \alpha_{\rm MLT})$\\
                        \rule{0pt}{10pt} $\left[\frac{{\rm Fe}}{{\rm H}}\right]=0.1$ dex model &1&0.7070 &0.2754&0.0176&1.8&15.25&198&$(M, \alpha_{\rm MLT})$\\
                        \rule{0pt}{10pt}$\alpha=1.65$ model &1&0.7155 &0.2703&0.0142&1.65&15.23&199&$(M, \alpha_{\rm MLT})$\\
                        \rule{0pt}{10pt}$\alpha=1.5$ model &1&0.7155 &0.2703&0.0142&1.5&15.23&201&$(M, \alpha_{\rm MLT})$\\
                        \rule{0pt}{10pt}$\alpha=1.7$ model &1&0.7155 &0.2703&0.0142&1.7&15.3&199&$(M, \alpha_{\rm MLT})$\\
                        \rule{0pt}{10pt}$\Delta\nu=10~\mu{\rm Hz}$ model&1&0.7155 &0.2703&0.0142&1.8&9.09&100&$M, (M, \alpha_{\rm MLT})$\\
                        \rule{0pt}{10pt}$M=1.3~M_\odot$ model &1.3&0.7155 &0.2703&0.0142&1.8&15.27&213&$M, (M, \alpha_{\rm MLT})$\\
                        \rule{0pt}{10pt}$M=1.7~M_\odot$ model &1.7&0.7155 &0.2703&0.0142&1.8&15.27&231&$M, (M, \alpha_{\rm MLT})$\\
                        \hline
                \end{tabular}
                \tablefoot{The [Fe/H] values are computed from the metallicities using the protosolar $(Z/X)_\odot=0.0199$ from \cite{asplund2009}.}
        \end{table*}
\section{Selection of reference models}
\label{secdipselect}

    In this work, we follow a grid-based approach to find reference models using the grids of stellar models described in Sect.~\ref{secstellarmodels}.  So far, only rotationally split dipole modes have been used for rotational inversions \citep[][see also Appendix.~\ref{secrotinvapp}]{deheuvels2012, deheuvels2014, dimauro2016, triana2017, deheuvels2017}. Hence, reference models need to be selected based on the properties of their dipole mode frequencies. As a metric for similar dipole mode properties between the observation and the reference model, we compute the Pearson correlation coefficient between the observed rotational splittings $\delta\omega$ and the reference model mode-trapping parameter $\zeta$:
    \begin{align}
        \rho=\frac{{\rm Cov} (\delta\omega,\zeta)}{\sigma_{\delta\omega}\sigma_\zeta}
    \label{eqpearson}
    \end{align}
     where $\zeta$ is computed as the ratio of the core to the total mode inertia  $I_{\rm core}/I_{\rm total}$. In the following, we refer to $\rho$ as the splitting correlation coefficient.

\begin{figure}
   \centering
   \includegraphics{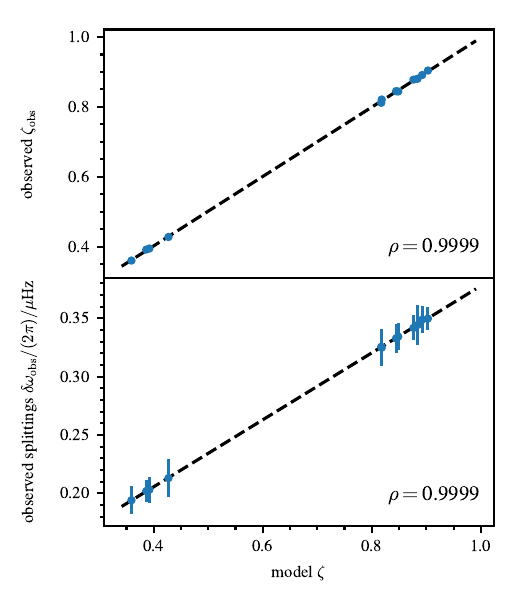}
      \caption{{\it Upper panel:}  Mode trapping parameter $\zeta_{\rm obs}$ of the synthetic star as a function of the mode trapping parameter $\zeta$ of the potential reference model. The black, dashed line is a fit to the data to illustrate the linear relation. {\it Lower panel:} Rotational splittings $\delta\omega_{\rm obs}$ of the synthetic star as a function of the mode trapping parameter $\zeta$ of the potential reference model.}
         \label{figcorrinertia}
   \end{figure}

     We illustrate the relation between the rotational splittings and the mode-trapping parameter in Fig.~\ref{figcorrinertia}, where we show the synthetic rotational splittings and the corresponding mode-trapping parameter $\zeta_{\rm obs}$ as a function of $\zeta$ of a reference model with a high correlation coefficient $\rho$. As shown by \cite{goupil2013}, the rotational splittings depend linearly on the mode-trapping parameter $\zeta$, which characterises the {\it p}/{\it g}-fraction of a mode. Clearly, the observed $\zeta_{\rm obs}$ correlate similarly well with the reference model $\zeta$ as the observed rotational splittings. A high correlation coefficient $\rho$ therefore ensures that the oscillation modes of the observed star and the reference model have almost the same {\it p}/{\it g}-fractions---a prerequisite for accurate rotational inversion results. We therefore suggest using this relation between the observed (synthetic) rotational splittings and $\zeta$ to find suitable reference models by only selecting models with a high enough splitting correlation coefficient for the rotation inversions.
     
    The calculation of dipole mode frequencies is computationally expensive and a grid based model fitting approach becomes impracticable very quickly. Hence, we apply two more steps before selecting models based on $\rho$. We first select models based on global seismic properties ($\Delta\nu, \nu_{\rm max}$) by setting a threshold for the maximum difference from the observed value (see Appendix~\ref{secglobselect}). For consistency, the global seismic parameters are again computed from the same scaling relations as for the synthetic data. In the second step, we impose a threshold on the $\chi^2$ of the radial modes ($\chi^2_{\rm rad}$) (see Appendix~\ref{secradselect}):
       \begin{equation}
        \chi^2 = \frac{1}{N} \sum_i \left( \frac{\nu_{i, {\rm obs}} - \nu_{i, {\rm mod}}}{\sigma_i} \right)^2
        \label{eqchi2_definition}
   \end{equation}
   where $\nu_{\rm obs}$ denotes the observed frequencies (synthetic or actual observation), $\nu_{\rm mod}$ denotes the frequencies of the potential reference stellar model and $\sigma_i$ denotes the uncertainties of the observed frequencies.

    For all stellar models that passed the first two steps, we compute dipole mode frequencies and rotational kernels following the procedure described in Sect.~\ref{secsynth}. As for the radial modes, we match the dipole modes of the models with the observed frequencies based on proximity in frequency domain.
    In addition to $\rho$, we also constrained the set of reference models using $\chi^2_{\rm dip}$ computed as in Eq.~(\ref{eqchi2_definition}). While using $\chi^2_{\rm dip}$ ensures similar frequencies, using $\rho$ ensures similar characters of the modes in the reference model and the observation. We assess the distribution of $\rho$ and the dipole-mode $\chi^2_{\rm dip}$ values for different reference models and the relation between the inversion results and the two metrics in the next section. The value of $\chi^2$ does of course depend on the uncertainty values, but also on the absolute values of the frequencies, which makes the $\chi^2$ dependent on the evolutionary state of the star. This makes it very difficult to give a universal threshold of $\chi^2$ that ensures a good fit to the observations. To take this into account the threshold value is chosen empirically (see discussion in Appendix~\ref{secthresh}). 

\section{Estimating internal rotation rates}\label{secensemble}

We now describe how we used the reference models to compute an estimate of the internal rotation rates. We also quantify the uncertainty introduced due to structural differences between the observed star and the reference models. 
\begin{figure*}
   \centering
   \includegraphics{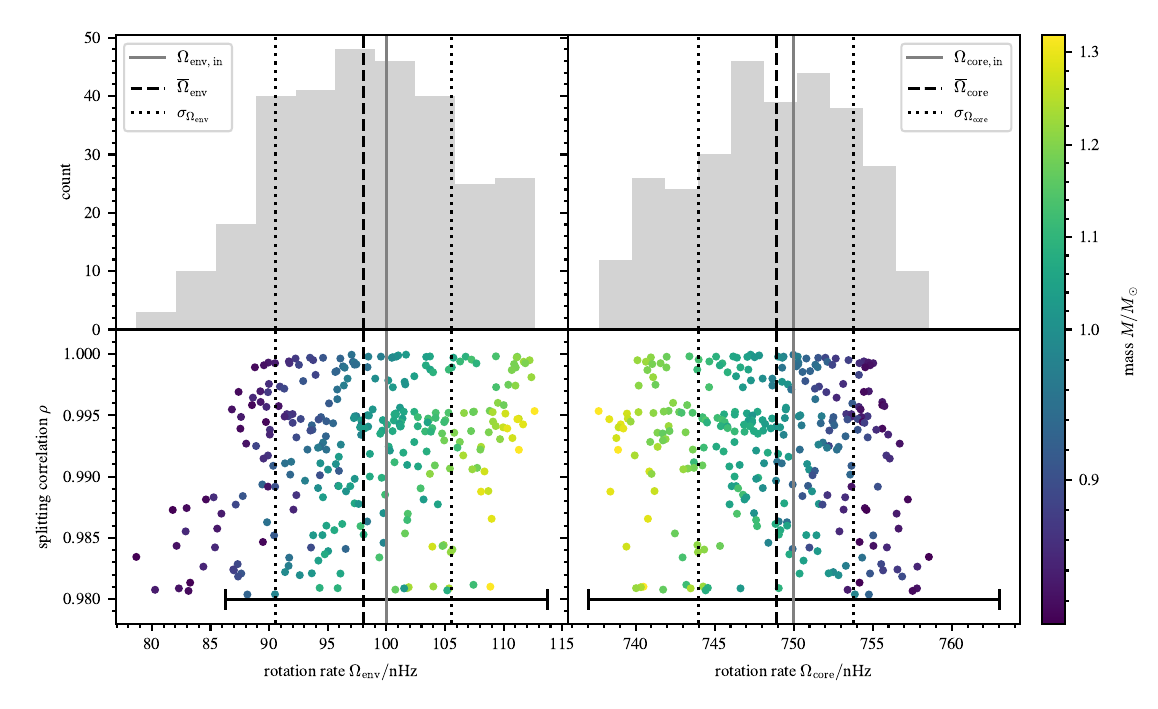}
      \caption{{\it Upper panels:} Histogram of the estimated envelope and core rotation rates in the left and right panel, respectively. The grey line indicates the input value used to compute the synthetic rotational splittings, the black dashed line indicates the mean value, and the black dotted line indicates $\pm$ one standard deviation from the mean. {\it Lower panels:} splitting correlation $\rho$ as a function of the estimated envelope and core rotation rate. A threshold of $\rho_{\rm thresh}=0.98$ was used here. The vertical lines have the same meaning as in the upper panel. The error bar indicates the maximal random error across all rotational inversion results selected.}
         \label{figensemblesurf}
   \end{figure*}

 As a first test case, we selected reference models and computed rotational inversions given the synthetic observables of our `fiducial model' (see first row of Table~\ref{tabsynobs} and Sect.~\ref{secsynth}). In the left panels of Fig.~\ref{figensemblesurf} we show the distribution of the splitting correlation coefficient as a function of the estimated envelope rotation rates, for all models with $\rho\geq0.98$, as well as the corresponding histogram. The same is shown for the estimated core rotation rates in the right panels. Using the median value of these distributions we estimated the core and envelope rotation rates:
\begin{align*}
    \overline{\Omega}_{\rm core}&=(749\pm13_{\rm rand}\pm5_{\rm ref})~{\rm nHz}\\
    \overline{\Omega}_{\rm env}&=(98\pm14_{\rm rand}\pm8_{\rm ref})~{\rm nHz}
\end{align*}
where we used target radii of $0.003~R$ and $0.98~R$ for the core and envelope rotation rate, respectively, and an error suppression parameter of $\mu=0$ (see also second row of Table~\ref{tabresults}). The comparison to the input values of $\Omega_{\rm core}=750$~nHz and $\Omega_{\rm env}=100$~nHz (see also first row of Table~\ref{tabresults}) shows that the inputs are recovered within the uncertainties. In the following, we refer to this estimation as the `ensemble inversion method'.

The deviations from the input rotation rate visible in Fig.~\ref{figensemblesurf} arise predominantly due to the mismatch between the chosen reference models and the structure of the observed star, since the uncertainties due to the inversion method have been shown to be negligibly small \citep[][their Fig.~6]{ahlborn2022}. We refer to these deviations as `systematic errors'. We note that other inversion methods (e.g. MOLA or SOLA) could be used to invert for the internal rotation rates for each reference model. In our work we have focussed on using eMOLA as it has been shown to work better for red-giant envelope rotation rates \citep{ahlborn2022}. For the core rotation rates, we find equivalent estimates for all three methods. For each reference model, the rotational inversion also provides a random error propagated from the uncertainties of the rotational splittings (see Eq.~(\ref{eqobjeMOLA}) and Appendix~\ref{secrotinvapp}). As a random error of the ensemble estimate (indicated with `rand') we give the maximum random error found in the ensemble of inversions calculated with all models that got selected based on the threshold in $\rho$. To compute the overall uncertainty of the ensemble inversion result introduced by discrepant structures of the reference models, we compute the standard deviation of the distributions shown in Fig.~\ref{figensemblesurf}. We refer to this uncertainty as `reference model uncertainty' (indicated with `ref'). For the distributions shown in Fig.~\ref{figensemblesurf} this reference model uncertainty amounts to 5 and 8~nHz for the core and envelope rotation rate, respectively, comparable to the random errors. This shows that for the given threshold values of $\rho$ and $\chi^2_{\rm dip}$ the impact of the discrepant reference model structure is on the same order of magnitude as the random errors. As the systematic errors of the envelope rotation rates increase relatively quickly with decreasing $\rho$, we set a more conservative threshold of $\rho_{\rm thresh}=0.98$ below which estimates are discarded. For the dipole-mode $\chi^2_{\rm dip}$ we chose again a rather large threshold value of 500 for the fiducial model, above which reference models are discarded. The selection of an appropriate threshold value is discussed in Appendix~\ref{secthresh}. We note that in contrast to the $\chi^2_{\rm dip}$ the splitting correlation coefficient always covers the same range of values between zero and one, which allows keeping the same threshold value for different synthetic observations.

The estimated envelope rotation rates, shown in the lower left panel of Fig.~\ref{figensemblesurf}, show a clear positive correlation with the reference model mass. This correlation can be explained by looking at sensitivities of the individual modes. We find that for the models more massive than the model of the synthetic observation, the {\it p}-dominated modes become more {\it p}-dominated while the sensitivities of the {\it g}-dominated modes stay approximately the same. The opposite applies for models with masses lower than the model of the synthetic observation. This means that in the inversion process, the kernels of the more massive models are matched with rotational splittings that are too large as compared to the sensitivity of the kernels. This leads to an overestimation of the envelope rotation and likewise an underestimation of the core rotation rate (see right panels of Fig.~\ref{figensemblesurf}). Again, the opposite argumentation applies to models with masses lower than the mass of the synthetic observation.

In addition to selecting models from the $M,\alpha_{\rm MLT}$-grid, we repeated the procedure with reference models selected from the smaller $M$-grid. The result is shown in the third row of Table~\ref{tabresults}. Despite small variations in the final estimates, the core and envelope rotation rates are recovered within the random errors. The standard deviation of the distribution, given as the reference model uncertainty, is likewise varying insignificantly compared to the larger $M,\alpha_{\rm MLT}$-grid. As long as the number of models is large enough to obtain a well constrained mean value and no bias is introduced, the reference model uncertainty does not seem to depend on the grid size (see discussion in Sect.~\ref{sectgridsize} and Appendix~\ref{secthresh}). We therefore conclude that the standard deviation of the distribution shown in Fig.~\ref{figensemblesurf} may be interpreted as the uncertainty due to structural differences between the observed star and the reference model. In conclusion, we find an unbiased estimate of the core and envelope rotation rates, recovering the input values within the uncertainties for the fiducial model. 
  \begin{figure*}
   \centering
   \includegraphics{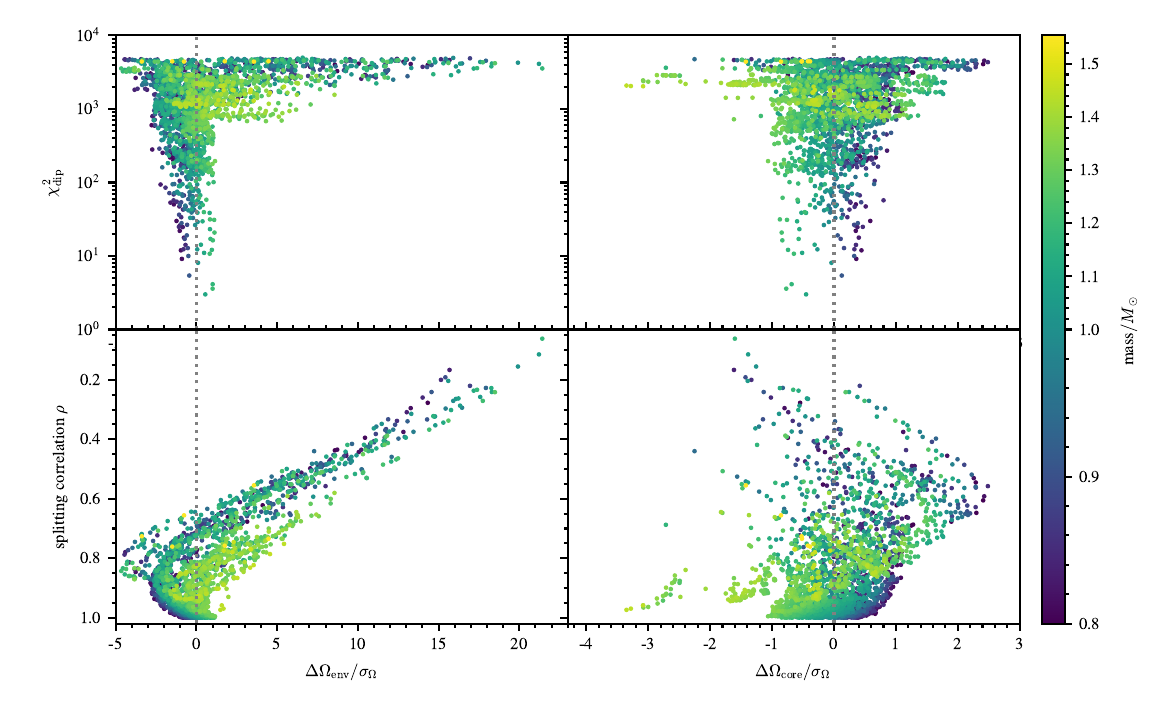}
      \caption{{\it Upper panels:} Dipole-mode $\chi_{\rm dip}^2$ as a function of the systematic error of the envelope and core rotation rate measured in units of the standard deviation in the left and right panel, respectively. {\it Lower panels:} Same as upper panels for the splitting correlation coefficient $\rho$. The initial mass of the reference model is colour coded. Note that the y-axis increases downwards.}
         \label{figomegachi2}
   \end{figure*}
   
In the following, we compare the splitting correlation coefficient and $\chi^2_{\rm dip}$ as metrics. In the left panels of Fig.~\ref{figomegachi2} we show the resulting dipole-mode $\chi^2_{\rm dip}$ and the splitting correlation coefficient as a function of the scaled systematic error of the envelope rotation rate, respectively. We find increasing systematic errors for increasing values of $\chi^2_{\rm dip}$ and decreasing values of $\rho$, both indicating a decreasing agreement between the synthetic observation and the reference model. While there are no large systematic errors for low $\chi^2_{\rm dip}$ values, we do find small systematic errors for large values of $\chi^2_{\rm dip}$. This indicates that the dipole-mode $\chi^2_{\rm dip}$ is not a unique measure for the suitability of the reference model for the rotational inversions. Using $\rho$ as a metric, we find fewer models with a low correlation and a low systematic error. For both metrics the inferred envelope rotation rates scatter around the input value even for high values of $\rho$ and low values of the dipole mode $\chi^2_{\rm dip}$. This implies that picking a single best fit reference model can lead to large deviations in the estimated rotation rates. The discussion of the ensemble inversion method above shows how an ensemble of reference models can mitigate this scatter and lead to an unbiased estimate of the underlying rotation rate.

In the right panels of Fig.~\ref{figomegachi2} we show the dipole-mode $\chi^2_{\rm dip}$ and the splitting correlation coefficient $\rho$ as a function of the systematic error of the core rotation rate measured in units of the standard deviation. As for the envelope rotation rate in the left panels of Fig.~\ref{figomegachi2}, we find increasing systematic errors for increasing values of $\chi^2_{\rm dip}$ and decreasing values of $\rho$. For the core rotation rate, the systematic errors do not exceed $2\sigma$ for the majority of the reference models, which makes it better constrained than the envelope rotation rate. This is already expected from previous studies \citep[e.g.][]{deheuvels2012,dimauro2016}. We further note that at least for the envelope rotation the correlation between the systematic errors and the splitting correlation coefficient $\rho$ is higher than for the dipole mode $\chi^2_{\rm dip}$. We therefore consider the splitting correlation coefficient to be a better metric to identify suitable reference models for rotational inversions.
\section{Impact of a surface perturbation}
\label{secsurfpertrot}
In the previous section, we used reference models that have the same surface layers as the models used to compute the synthetic observations. It is, however, well known that most state of the art stellar evolution codes do not model the surface layers of stars accurately. This leads to the previously described surface effect. This surface effect only acts on the pure {\it p}-mode component of a mixed mode. Hence, a different near surface structure leads to the coupling of a different pure {\it p}- and {\it g}-mode to form a different mixed mode. This does also change the sensitivity kernels of the rotational splittings, necessitating a proper discussion of the surface effect in terms of rotational inversions \citep{ong2024} when using a single reference model for the inversion. Here, we investigate the impact of a surface perturbation on rotational inversion results when using an ensemble of reference models selected based on matching the character of the observed modes with the modes of the reference model.

To study the impact of a surface perturbation on the rotational inversion results, we applied the ensemble inversion method to the surface perturbed model described in Sect.~\ref{secsurfpert}. In a first step, we selected references models from the stellar model grid without applying any surface correction to the oscillation mode frequencies. We find the following core and envelope rotation rates:
\begin{align*}
    \overline{\Omega}_{\rm core}&=(747\pm 14_{\rm rand}\pm 8_{\rm ref})~{\rm nHz}\\
    \overline{\Omega}_{\rm env}&=(105\pm 17_{\rm rand}\pm 11_{\rm ref})~{\rm nHz}
\end{align*}
This shows that despite the difference in the near surface layers between the synthetic observation and the reference models, the ensemble inversion is still able to recover the input rotation rate within the uncertainties. At first glance, this might be surprising. However, by construction, the ensemble inversion method does only select reference models that have a high correlation between the observed rotational splittings and the reference model mode trapping parameter $\zeta$. This ensures that the sensitivity to rotation in the modes of the reference models is very close to the sensitivity of the modes in the observed star, which is a necessity to recover accurate rotation results. This highlights the importance of reproducing the mixed mode character when selecting reference models for rotational inversions.

While correcting for the surface effect remains very difficult for mixed modes, it is possible to obtain surface corrected {\it p}-mode frequencies \citep[e.g.][]{ball2014}. As the radial modes are pure {\it p}-modes, their frequencies can be corrected for the surface effect. We apply the \cite{ball2014} two term surface correction to correct the radial mode frequencies of our reference models. The two parameters of the \cite{ball2014} correction are fit for each reference model individually. For the selection of the reference models, we follow the same procedure as described in Sect.~\ref{secdipselect}. Here, we compare the $\Delta\nu$ obtained from a linear fit of the radial mode frequencies of the synthetic observations to the scaling relation value of $\Delta\nu$ of the reference models. We apply a threshold of 10 for the radial mode $\chi^2_{\rm rad}$. We note that this is much lower than the threshold value applied for frequencies without surface term correction. This occurs as the model by model fit of the \cite{ball2014} parameters does also remove frequency differences that are not due to the surface effect. For the reference models selected as described above, we obtain the following ensemble inversion results:
\begin{align*}
    \overline{\Omega}_{\rm core}&=(748\pm14_{\rm rand}\pm6_{\rm ref})~{\rm nHz}\\
    \overline{\Omega}_{\rm env}&=(103\pm16_{\rm rand}\pm9_{\rm ref})~{\rm nHz}
\end{align*}
As for the previous case, we are able to recover the input rotation rates when correcting the radial mode frequencies for the surface effect.

\section{Dependence on stellar parameters}
\label{secparamdep}
In this section, we explore how the rotation rates estimated with the ensemble inversion method depend on different fundamental stellar parameters like mass, initial composition, mixing length parameter and different positions along the RGB. To this end, we systematically vary each parameter separately in the following subsections. The estimated rotation rates for all synthetic observations and different grids are summarised in Table~\ref{tabresults} and visualised in Fig.~\ref{figsummary}. The comparison with the input rotation rates shows that we are able to recover the underlying rotation rates in all but one of the synthetic observations. The different synthetic observations used in this section are summarised in Table~\ref{tabsynobs}. We compute the total error of our estimate by means of error propagation:
\begin{align*}
\sigma_{\rm total}=\sqrt{\sigma_{\rm rand}^2+\sigma_{\rm ref}^2}
\end{align*}
where $\sigma_{\rm rand}$ refers to the random uncertainty and $\sigma_{\rm ref}$ to the reference model uncertainty of the final estimate. In the remainder of this section, we select the reference models using both the splitting correlation coefficient and the dipole-mode $\chi^2_{\rm dip}$. A comparison of the results when using either of them alone is shown in Appendix~\ref{secmetricsapp}. The threshold values for $\chi^2_{\rm dip}$ are determined as described in Appendix~\ref{secthresh}.
\begin{table*}
	\caption{Rotational inversion results for different synthetic observations using $\rho$ and $\chi^2_{\rm dip}$ as metrics simultaneously.}
	\label{tabresults}
	\centering
	\renewcommand{\arraystretch}{1.2}
	\begin{tabular}{c c c c c c c c c c}
		\hline\hline
		\rule{0pt}{12pt}name&$\Omega_{\rm core}$&$\sigma_{\rm rand}$&$\sigma_{\rm ref}$&$\Omega_{\rm env}$&$\sigma_{\rm rand}$&$\sigma_{\rm ref}$&$\rho_{\rm thresh}$&$\chi^2_{\rm dip, thresh}$&grid\\
		\hline
		\rule{0pt}{10pt}input&750&--&--&100&--&--&--&--&--\\
		\hline
		\rule{0pt}{10pt}$M=1~M_\odot$&749&13& 5&98&14& 8&0.98&500&$M, \alpha_{\rm MLT}$\\
		\hline
		\rule{0pt}{10pt}$M=1~M_\odot$&752&14& 4&91&15& 8&0.95&500&$M$\\
		\hline
		\rule{0pt}{10pt}$\left[{\rm Fe}/{\rm H}\right]=-0.1$ dex&744&14& 7&105&16&11&0.98&500&$M, \alpha_{\rm MLT}$\\
		\hline
		\rule{0pt}{10pt}$\left[{\rm Fe}/{\rm H}\right]=+0.1$ dex&748&14& 6&103&14&13&0.98&500&$M, \alpha_{\rm MLT}$\\
		\hline
		\rule{0pt}{10pt}$\Delta\nu=9~\mu{\rm Hz}$&746&17& 4&107&22& 7&0.98&100&$M, \alpha_{\rm MLT}$\\
		\hline
		\rule{0pt}{10pt}$\Delta\nu=9~\mu{\rm Hz}$&745&17& 4&112&22&11&0.95&100&$M$\\
		\hline
		\rule{0pt}{10pt}$\alpha_{\rm MLT}=1.5$&751&14& 3&93&22& 7&0.98&500&$M, \alpha_{\rm MLT}$\\
		\hline
		\rule{0pt}{10pt}$\alpha_{\rm MLT}=1.65$&746&15& 8&107&18&12&0.98&500&$M, \alpha_{\rm MLT}$\\
		\hline
		\rule{0pt}{10pt}$\alpha_{\rm MLT}=1.7$&746&15& 7&101&16&11&0.98&500&$M, \alpha_{\rm MLT}$\\
		\hline
		\rule{0pt}{10pt}$M=1.3~M_\odot$&751&13& 4&94&15&12&0.95&500&$M$\\
		\hline
		\rule{0pt}{10pt}$M=1.3~M_\odot$&750&13& 4&95&14&10&0.98&500&$M, \alpha_{\rm MLT}$\\
		\hline
		\rule{0pt}{10pt}$M=1.7~M_\odot$&751&13& 7&129&13& 8&0.95&1000&$M$\\
		\hline
		\rule{0pt}{10pt}$M=1.7~M_\odot$&751&12& 5&127&12& 7&0.98&1000&$M, \alpha_{\rm MLT}$\\
		\hline
		\rule{0pt}{10pt}surf. pert.&747&14& 8&105&17&11&0.98&500&$M, \alpha_{\rm MLT}$\\
		\hline
		\rule{0pt}{10pt}surf. pert., surf. corr.&748&14& 6&103&16& 9&0.98&500&$M, \alpha_{\rm MLT}$\\
		\hline
	\end{tabular}
	\tablefoot{The rotational inversion results are computed with the ensemble rotational inversion described in Sect.~\ref{secensemble}. All rotation rates and uncertainties given in units of nHz.}
\end{table*}
      
  \begin{figure}
   \centering
   \includegraphics{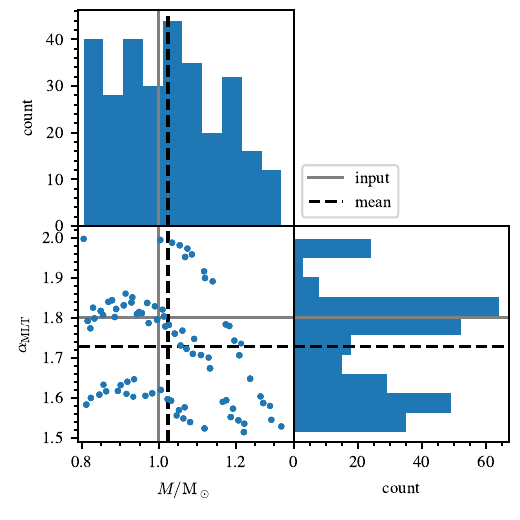}
      \caption{{\it Upper left:} Distribution of initial stellar masses for selected reference models. {\it Lower right:} Distribution of mixing length parameters for selected models. {\it Lower left:} Scatter plot showing the mixing length parameters versus the initial stellar mass for selected stellar models. The stellar models were selected following the selection process described in Sects.~\ref{secdipselect} and \ref{secensemble} using the fiducial model. The models were selected from the $M, \alpha_{\rm MLT}$-grid. The black dashed lines and the grey lines refer to the actual mean and the input value, respectively.}
         \label{figalphamassfinal}
   \end{figure}
\subsection{The effect of the initial mass}
\label{secstellarmass}
For most problems in stellar physics, the stellar mass is the most important parameter. To study the dependence on the initial stellar mass we estimate the rotation rates of the fiducial model, the $M=1.3~M_\odot$ model and the $M=1.7~M_\odot$ model (see Table~\ref{tabsynobs} for the stellar parameters) by selecting reference models from the $M, \alpha_{\rm MLT}$-grid. The computations with different stellar masses show that the results of the ensemble inversions do not depend strongly on the mass of the observed star. In conclusion, an accurate mass estimate is not necessary to accurately recover the rotation rates, and conversely, an accurate mass does not necessarily recover the rotation rates accurately.

We show in Fig.~\ref{figensemblesurf} for the fiducial model that models with a larger range of masses are able to recover the input rotation rates. The corresponding histogram of the stellar masses in the set of reference models is shown in Fig.~\ref{figalphamassfinal}. Reference stellar models for the ensemble inversion were selected in a range of 0.8 to 1.4~M$_\odot$ with a mean value $\overline{M}\approx1.02~{\rm M}_\odot$ close to the input value. The scatter plot in the lower left panel shows the mixing-length parameter as a function of the stellar mass. Clearly, this parameter space is not populated uniformly, instead a pattern of gaps shows up. This pattern arises due to the selection of models based on the splitting correlation coefficient $\rho$ and $\chi^2_{\rm dip}$. The emergence of the pattern shows that only certain combinations of $M$ and $\alpha_{\rm MLT}$ result in a reference model suitable for the rotational inversion. The spread in values of the mass and the mixing-length parameter indicates that a deviation in one parameter may be compensated by a deviation in the other one.
Before applying the cut in $\rho$ and $\chi^2_{\rm dip}$ the $M,\alpha_{\rm MLT}$ parameter space is more uniformly populated and the mass of the input model is reproduced by the mean value of the mass distribution (see blue results in Fig.~\ref{figalphamassinterm}, upper left panel). The selection of suitable reference models based on the splitting correlation coefficient shows that reproducing the global seismic properties and the radial modes of the observed stars is not sufficient to reproduce the rotation rates. We find similar results when using the $M$-grid, where we keep a fixed value of $\alpha_{\rm MLT}=1.8$. Due to the smaller number of models in the $M$-grid less suitable models can be found for the ensemble inversion method. To increase the number of selected models we lowered the threshold value to $\rho_{\rm thresh}=0.95$ at the price of a slightly reduced accuracy.

For the $M=1.3~M_\odot$ model, we can confirm the result found for the fiducial model and recover the input rotation rates within the random uncertainties while using a range of masses in the set of reference models independent of the grid where the reference models were chosen from.

For the $M=1.7~M_\odot$ model, the results need to be interpreted with more care. While we are able to recover the core rotation rate within the random uncertainties, this is not possible for the envelope rotation rate. A closer investigation shows, however, that it is not possible to recover the envelope rotation even when using the stellar model that was used to create the synthetic data as a reference model. In the stellar model used to create the synthetic observations, the boundary of the fast rotating core reaches into the {\it p}-mode cavity. However, eMOLA suppresses the sensitivity of the surface averaging kernel only below the {\it p}-mode cavity. Hence, there is residual sensitivity to the fast core rotation, increasing the estimate of the envelope rotation. When assuming that the step is located deeper inside the star, for example at 1.5 times the radius of the hydrogen burning shell, the sensitivity of the surface averaging kernel does no longer reach into the fast rotating core. In such a case we are again able to recover the input rotation. It is of course unknown a~priori which situation prevails in a star. We note that the original MOLA inversions are subject to the same behaviour. We conclude that the mismatch of the envelope rotation for the $M=1.7~M_\odot$ model with the input values is not due to the ensemble inversion, but rather due to the inversion method itself in combination with the envelope step rotation profile \citep[see discussion in][Sect. 3.3]{ahlborn2022}. These results are again independent of the stellar model grid used.

\subsection{The effect of initial composition}
Another parameter influencing the structure and evolution of stars is the initial chemical composition. We have constructed two synthetic observations with metallicity values of $[{\rm Fe}/{\rm H}]=\pm 0.1$~dex (rows two and three in Table~\ref{tabsynobs}), what is on the same order as typical uncertainties on observed metallicities. We selected reference models from the $M, \alpha_{\rm MLT}$-grid. We find that both the core and the envelope rotation rate can be recovered within the derived uncertainties (see Table~\ref{tabresults}, rows four and five). This is possible despite the fact that the grid used has only a single chemical composition, that is, the solar composition. We note that the discrepancy between the estimated and the input rotation rates increases for an increased mismatch in the chemical composition. Therefore, we conclude that the metallicity of the reference model needs to be constrained to within $\pm0.1$~dex to obtain reliable inversion results. 

For our stellar model grids and the synthetic observations, we have used the \cite{asplund2009} solar mixture of heavy elements (see Appendix~\ref{secstellarmodelsapp} and \ref{secsynthapp}). The composition of the Sun is a long-standing problem, however, and other compositions have been proposed in the literature. To test the impact of the solar composition on our results, we have created a synthetic observation very similar to the fiducial model, however, with the standard solar composition of \cite{grevesse1998} (GS98, $X=0.7062, Y=0.275, Z=0.0188$). We find the following rotation rates: 
\begin{align*}
    \overline{\Omega}_{\rm core}&=(748\pm16_{\rm rand}\pm7_{\rm ref})~{\rm nHz}\\
    \overline{\Omega}_{\rm env}&=(104\pm20_{\rm rand}\pm14_{\rm ref})~{\rm nHz}
\end{align*}
when applying the ensemble inversion method. Core and envelope rotation rates are recovered within the random uncertainties. This result is very comparable to the result for the [Fe/H]=+0.1~dex model (see Table~\ref{tabresults}). Given the  composition of the [Fe/H]=+0.1~dex model (see third row of Table~\ref{tabsynobs}), one would have expected a similar impact of the \cite{grevesse1998} composition on the result. Using the \cite{asplund2009} Z/X=0.0199 the GS98 model has a [Fe/H]$\approx0.13$~dex, slightly larger than the value we have tested previously. Based on experience from modelling the Sun we expect the effect to be smaller when only changing the relative distribution of heavy elements instead of changing the overall composition. We hence conclude that the choice of the solar composition does not have a strong impact on the final inversion result.
\subsection{The effect of the mixing length parameter}\label{secml}
The mixing-length parameter of the selected models, shown in Fig.~\ref{figalphamassfinal}, spans a range from 1.5 to 2, that is, the whole range of possible values on the grid. The gaps in the distribution of the mixing-length parameters are discussed already in Sect.~\ref{secstellarmass}. We find that we are able to recover the input rotation rates for all synthetic observations constructed with different input values of the mixing-length parameter. This is in agreement with our conclusion on the stellar mass that a range of values is suitable to recover the input rotation rate. We note that for the model with $\alpha_{\rm MLT}=1.5$ less models were available for the ensemble inversion as the $M,\alpha_{\rm MLT}$-grid does not extend below this value. This makes the estimated rotation rates for this synthetic observation statistically less robust.

The mixing length parameter is best constrained by the observed effective temperature. Changing the mixing-length parameter has a strong impact on the radius of stellar models of red giants, which in turn causes the effective temperature of the model to diverge from the observed value. We find however that constraining the effective temperature and as a consequence the mixing length parameter does not strongly impact on the final rotational inversion result. Therefore, we did not include $T_{\rm eff}$ as a constraint in the search for the reference models. For more details we refer to Appendix~\ref{secteffapp}
\subsection{Evolution along the RGB}
In addition to the initial stellar parameters, the position of the stars on the RGB is also expected to play a role in determining the internal rotation rates. To test this dependency, we evolved the fiducial model, discussed in Sect.~\ref{secensemble}, further up the RGB to a large frequency separation of $\Delta\nu=9.1~\mu$Hz. The eMOLA inversion method used here was formulated to estimate envelope rotation rates that do not show systematic errors for more evolved stars, assuming a known reference model \citep[see][]{ahlborn2022}. In Table~\ref{tabresults} and Fig.~\ref{figsummary} we show that the ensemble inversion method does recover the input rotation rates within the random uncertainty also for more evolved models. The reference model uncertainties $\sigma_{\rm ref}$ determined from the standard deviation in the distribution of the estimated rotation rates are very comparable to the fiducial model. We note however that the envelope rotation rates become biased to higher rotation values, while the core rotation rates becomes biased to lower rotation rates. Furthermore, the range of masses in the set of reference models is smaller, which indicates that a better constraint on the mass is needed for the more evolved models. These results change only marginally when selecting reference models from the $M$-grid, and the above conclusions remain unaffected.

\section{Discussion}
\subsection{Impact of the uncertainty model}
\label{secuncertdiss}
In contrast to the synthetic observables, the rotational splittings and frequencies observed in actual stars are subject to observational uncertainties. As described in Sect.~\ref{secsynth}, we describe the measurement uncertainties for the synthetic observables with an uncertainty model. Given the uncertainty model, we can determine the impact of these measurement uncertainties on the rotational inversion results. To test the impact of the frequency uncertainties, we have perturbed the radial and dipole mode frequencies of our synthetic mode set with normally distributed random numbers with a standard deviation given by the uncertainty model. The rotational splittings remained unperturbed as before. We find that always a very similar set of reference models gets selected for a given synthetic observation. Therefore, the ensemble inversion results do only vary marginally. We can hence conclude that the results of our method are not strongly influenced by the measurement uncertainties of the oscillation frequencies, given our uncertainty model.

Because we indirectly select the reference models based on the observed rotational splittings through the splitting correlation coefficient, the set of selected reference models could be potentially impacted by perturbations of the rotational splittings. To understand the behaviour of the ensemble inversions for perturbed rotational splittings we first assessed the behaviour of the individual rotational inversion results under perturbations of the rotational splittings. Following a Monte Carlo approach, we computed rotational inversions for multiple sets of perturbed rotational splittings. In each set, the rotational splittings were perturbed, with a normally distributed perturbation with zero mean and using the individual measurement uncertainties as the standard deviation. We verified that the random uncertainties of the individual inversion results, derived from error propagation, reflect the standard deviation of rotational inversion results computed for many realisations of perturbed rotational splittings.

\begin{figure}
    \centering
    \includegraphics{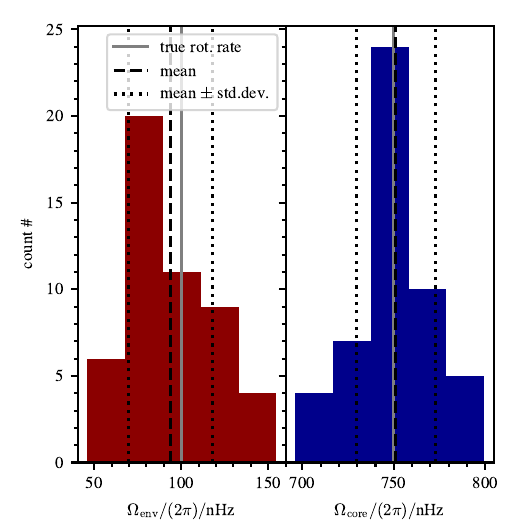}
    \caption{Ensemble inversion results for 50 perturbations of the rotational splittings within $2\sigma$. {\it Left panel:} Histogram of the envelope rotation rates estimated from ensemble inversions for different realisations of perturbed rotational splittings. The vertical dashed and dotted lines indicate the mean and the standard deviation, respectively. The vertical grey line indicates the input value. {\it Right panel:} Same as the left panel but for the core rotation rate.}
    \label{figenspert}
\end{figure}

Likewise, we tested the dependence of the ensemble inversion results on the uncertainties of the rotational splittings. Here, we leave the oscillation frequencies unperturbed. We repeated the Monte Carlo approach for the ensemble inversions, generating 50  sets of rotational splittings. We again perturbed the rotational splittings with normally distributed perturbations. However, in this case we used a standard deviation of $2\sigma$ to test the robustness of the ensemble inversion against larger perturbations of the rotational splittings.
As a direct consequence of the larger perturbations, we needed to use a lower threshold in the splitting correlation coefficient for the ensemble inversions as compared to the unperturbed case. However, the rotation rates estimated from individual reference models of the ensemble do not depend on the uncertainties as we set $\mu=0$. The histograms of the core and envelope rotation rates for 50 perturbations of the rotational splittings are shown in Fig.~\ref{figenspert}. We find that in both cases the mean value of the distribution recovers the input rotation rate within the uncertainties. Likewise, the standard deviation of the distribution reflects the random uncertainty of the individual rotational inversions. This shows that we can use the random uncertainties from the individual rotational inversions as an estimate for the random uncertainty of the ensemble inversion result. We conclude that the selection of reference models based on the splitting correlation coefficient and the final ensemble inversion estimate are robust against perturbations of the splittings.

The uncertainty model discussed in Appendix.~\ref{secuncert} does not differentiate between {\it p}-dominated or {\it g}-dominated dipole modes. However, the uncertainties are expected to depend on the mode character. As the {\it g}-dominated modes have a higher mode inertia, they appear with smaller width in the power spectra, leading to smaller uncertainties on the measured frequency values. To test the impact of this variation on our ensemble inversion results, we have recomputed our uncertainties by scaling them inversely with the mode inertia, such that {\it g}-dominated modes have smaller uncertainties than the {\it p}-dominated modes. We find that this does impact the results only marginally. Due to the smaller uncertainties, it is necessary to choose a larger threshold value in $\chi^2_{\rm dip}$ to be consistent with the results from the previous section. Additionally, the random uncertainties on the estimated core rotation rates decrease due to the smaller uncertainties in the {\it g}-dominated modes. Our conclusions remain otherwise unchanged.
\subsection{Impact of the mixed mode pattern}
\begin{figure}
    \centering
    \includegraphics{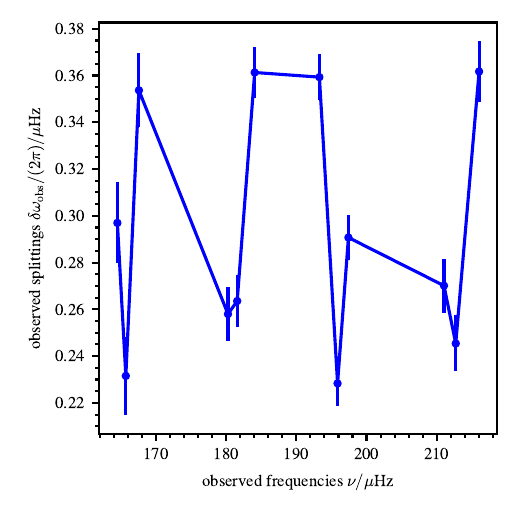}
    \caption{Rotational splittings as a function of frequency for a synthetic observation with $\alpha=1.7$ very similar to the model discussed in Sect.~\ref{secml}. In some acoustic radial orders no clear {\it p}-dominated mode is visible in this pattern, however, instead two somewhat {\it p}-dominated modes are present.}
    \label{figpattern}
\end{figure}

The ensemble inversion method is based on correlating the (synthetic) observed rotational splittings with the reference model mode trapping function $\zeta$. Hence, the observed mixed mode pattern has a direct impact on the selection of the reference models. In this section, we discuss the impact of this mixed mode pattern on the final results. To illustrate the impact we use the model with $\alpha=1.7$ described in Sect.~\ref{secml}. For the results presented in Table~\ref{tabresults} we used synthetic observations which showed one clearly {\it p}-dominated mode per acoustic radial order and otherwise g-dominated modes. For the $\alpha=1.7$ model we now discuss results for a pattern in which in some acoustic radial orders no clear {\it p}-dominated mode however instead two somewhat {\it p}-dominated modes are present. The rotational splittings as a function of frequency, illustrating the mixed mode pattern, are shown in Fig.~\ref{figpattern}.
 
When applying the ensemble inversion method with the threshold values $\rho=0.98$ and $\chi^2_{\rm dip}=500$ as in Sect.~\ref{secparamdep} we obtain:

\begin{align*}
    \overline{\Omega}_{\rm core}&=(747\pm16_{\rm rand}\pm5_{\rm ref})~{\rm nHz}\\
    \overline{\Omega}_{\rm env}&=(107\pm21_{\rm rand}\pm7_{\rm ref})~{\rm nHz}
\end{align*}
which is again reproducing the input rotation rate within the uncertainties. For a pattern with clear {\it p}-dominated modes and otherwise similar parameters (Table~\ref{tabresults}, $\alpha=1.7$), we were able to recover the input rotation rates with a smaller difference. This occurs primarily due to the mixed mode pattern in this synthetic observation shown in Fig.~\ref{figpattern}. The lack of a single, clearly {\it p}-dominated mode makes this pattern much more prone to erroneously matching {\it p}-dominated splittings with {\it g}-dominated kernels and vice versa. To improve the results of the ensemble inversion method for the pattern shown in Fig.~\ref{figpattern} further one could consider varying the threshold values for $\rho$ and $\chi^2_{\rm dip}$. As we do already recover the input rotation rate for the default parameters we do not consider this exercise necessary here.
\subsection{Impact of the rotation profile}\label{secrotcomp}
\begin{figure}
   \centering
   \includegraphics[width=\hsize]{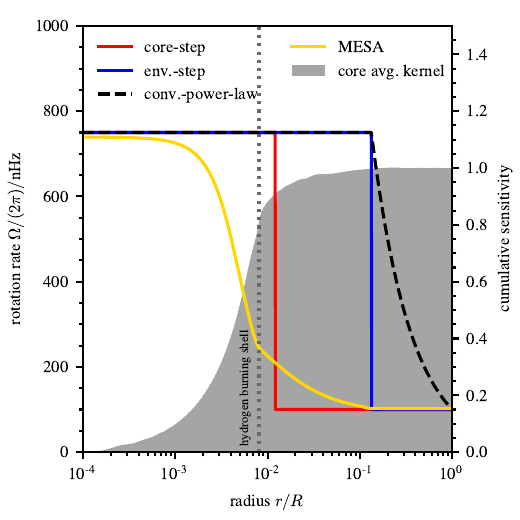}
      \caption{Rotation profiles as a function of fractional radius. Note that the radius is shown on a logarithmic scale. We show the three synthetic profiles used in this work in red, blue and black. The profile of the rotating MESA model is shown in yellow. We use $\Omega_{\rm core}/(2\pi)=750$~nHz and $\Omega_{\rm env}/(2\pi)=100$~nHz as default values for the synthetic rotation profiles. The vertical dotted line indicates the location of the hydrogen burning shell. The cumulative core averaging kernel of the fiducial model is shown with the grey shaded are.}
         \label{figrotprofiles}
   \end{figure}
In Sect.~\ref{secensemble} we used only one synthetic rotation profile for our calculations, i.e. the envelope step profile, featuring a step at the base of the convective envelope and otherwise constant rotation rates. However, in an observed star the internal rotation profile is a priori unknown. In this section, we therefore test how our ensemble inversion results depend on the underlying rotation profile, as the selection of reference models indirectly depends on the rotational splittings through the computation of the correlation coefficient. We generate the synthetic observables necessary for this test using the fiducial model and the synthetic rotation profiles described in Sect.~\ref{secsynthrot}. We summarise the results for the different synthetic rotation profiles in Table~\ref{tablerotcomp}. The core-step and convective power law profile are already used in \cite{ahlborn2022} for a single known reference model and eMOLA inversions. We show their results for comparison in Table~\ref{tablerotcomp} when applicable and demonstrate that the ensemble inversion results are in very good agreement with the results for a known reference model. This shows that the ensemble method is robust against variations of the shape of the rotation profile and recovers the best possible inversion result (obtained when the reference model is known).

Depending on the underlying rotation profile a mismatch of the estimated core rotation rate beyond the uncertainties may be present. The mismatch in case of the core step profile can be explained by the sensitivity of the individual rotational splittings. By inspecting the rotational kernels it is easy to see that about 10\% of the sensitivity of the {\it g}-component extends into the slow rotating region beyond 1.5$r_{\rm H}$. The resulting core averaging kernel is illustrated in Fig.~\ref{figrotprofiles}. At the location of the step in the core step profile, the core averaging kernel has gained approximately 90\% of its total sensitivity very similar to the {\it g} component of the individual kernels. Hence, the estimated core rotation rate which is an average value using the averaging kernel as the weighting function is sensitive to the slow rotating region and therefore has to be lower than the nominal core rotation rate. Given that the {\it g}-components of the different rotational kernels probe basically the same region \citep[][their Sect.~6.2]{ahlborn2022}, it is not possible to construct a core averaging kernel that is only sensitive to the region below 1.5$r_{\rm H}$. We note that we obtained nearly identical results for the core rotation rate when using MOLA or SOLA inversions instead of eMOLA. Finally, an MCMC could be used to determine the rotation rates and the location of the step simultaneously \citep{fellay2021}. Similarly, the surface averaging kernel is not exclusively sensitive to the surface layer of the star, but rather probes an average envelope rotation rate. In case of the convective power law profile, the rotation rate decreases throughout the envelope until it reaches its minimum at the surface. The estimate of the envelope rotation therefore has to be higher than the actual surface rotation rate as can be seen for the results shown in Table~\ref{tablerotcomp}.

We further computed ensemble inversions for the envelope step and core step profile with a varying contrast between the core and the envelope rotation. We have varied the ratio from a value of one (i.e. solid body rotation) up to 20 (the fiducial case has a ratio of 7.5). We have kept the envelope rotation rate at the value of 50 or 100~nHz. We summarise the results in the lower part of Table~\ref{tablerotcomp}. For a ratio of unity, the splittings of the {\it p}-dominated modes are larger than the splittings of the {\it g}-dominated modes due to the larger total integral of the rotational kernels. In this way, the splittings remain correlated with the mode trapping, even though with a negative correlation coefficient. We note that for a ratio of 2 the synthetic rotational splittings are the same for all modes within uncertainties. In this case there is no correlation between the splittings and the mode trapping. Nevertheless, the inversions are able to recover the input value correctly as the matching of observed splittings to modes from the reference model does not matter in the case of constant splittings. In the cases with a higher core-to-envelope ratio the method does work in the same way and with the same level of accuracy as for the fiducial case. We find, however, that the reference model uncertainty depends on the core-to-envelope ratio. It is lowest for a ratio of 2 (essentially constant rotational splittings). With increasing difference between the {\it p}- and {\it g}-dominated splittings the reference model uncertainty increases. In case of matching a mode with the wrong character of an observed splitting the error in the estimated rotation rate increases with the difference between the {\it p}- and {\it g}-dominated splittings. We conclude that the ensemble inversion method is robust against variations of the core to envelope ratio.
        \begin{table*}
                \caption{Comparison of core and envelope rotation rates obtained with the ensemble inversions and different synthetic rotation profiles}
                \label{tablerotcomp}
                \centering
                \begin{tabular}{c c c c c c c c}
                        \hline\hline
                        Profile&Method&$\Omega_{\rm core}$&$\sigma_{\rm rand}$&$\sigma_{\rm ref}$&$\Omega_{\rm env}$&$\sigma_{\rm rand}$&$\sigma_{\rm ref}$\\
                        \hline
                        \rule{0pt}{10pt}Core step &ensemble&$690$&13&5&100&14&7\\
                        &\hphantom{e}eMOLA&$690$ &6&--&$102$&6&--\\
                        \hline
                        \rule{0pt}{10pt}Envelope step&ensemble&$749$&13&5&$98$&14&8\\
                        &\hphantom{e}eMOLA & $749$ &6&--&$100$&6&--\\
                        \hline
                        \rule{0pt}{10pt}Convective power law&ensemble&$750$&13&4&$153$&14&6\\
                        &\hphantom{e}eMOLA& $750$ &6&--&$152$&6&--\\
                        \hline\hline
                        \rule{0pt}{10pt}Envelope step,&&&&&&&\\
                        \rule{0pt}{10pt}$\Omega_{\rm env}=100$~nHz&&&&&&&\\
                        \hline
                        \rule{0pt}{10pt}$\Omega_{\rm core}/\Omega_{\rm env}=1$&ensemble& 100 &14&1&100&14&2\\
                        \rule{0pt}{10pt}$\Omega_{\rm core}/\Omega_{\rm env}=2$&ensemble& 200 &14&1&100&18&1\\
                        \rule{0pt}{10pt}$\Omega_{\rm core}/\Omega_{\rm env}=3$&ensemble& 300 &13&1&100&14&2\\
                        \rule{0pt}{10pt}$\Omega_{\rm core}/\Omega_{\rm env}=5$&ensemble& 500 &13&3&99 &14&5\\
                        \rule{0pt}{10pt}$\Omega_{\rm core}/\Omega_{\rm env}=10$&ensemble& 998 &13&8&97 &14&11\\
                        \hline\hline
                        \rule{0pt}{10pt}Core step,&&&&&&&\\
                        \rule{0pt}{10pt}$\Omega_{\rm core}/\Omega_{\rm env}=20$&&&&&&&\\
                        \hline
                        \rule{0pt}{10pt}$\Omega_{\rm env}=100$~nHz&ensemble& 1826 &13&15&99 &14&23\\
                        \rule{0pt}{10pt}$\Omega_{\rm env}=50$~nHz&ensemble& 913 &13&8&50 &14&12\\
                        \hline
                \end{tabular}
                \tablefoot{The profile type used to compute the synthetic rotational splittings is given in the first column. Core and envelope rotation rates have been calculated with target radii of $r_0/R=0.003$ and $r_0/R=0.98,$ respectively. When varying the core-to-envelope ratio the envelope rotation rate was set to 50 or a 100~nHz. The eMOLA results are taken from \citealt{ahlborn2022} (their Table~1).}
        \end{table*}

\subsection{Impact of the grid properties}\label{sectgridsize}
\begin{figure*}
    \centering
    \includegraphics{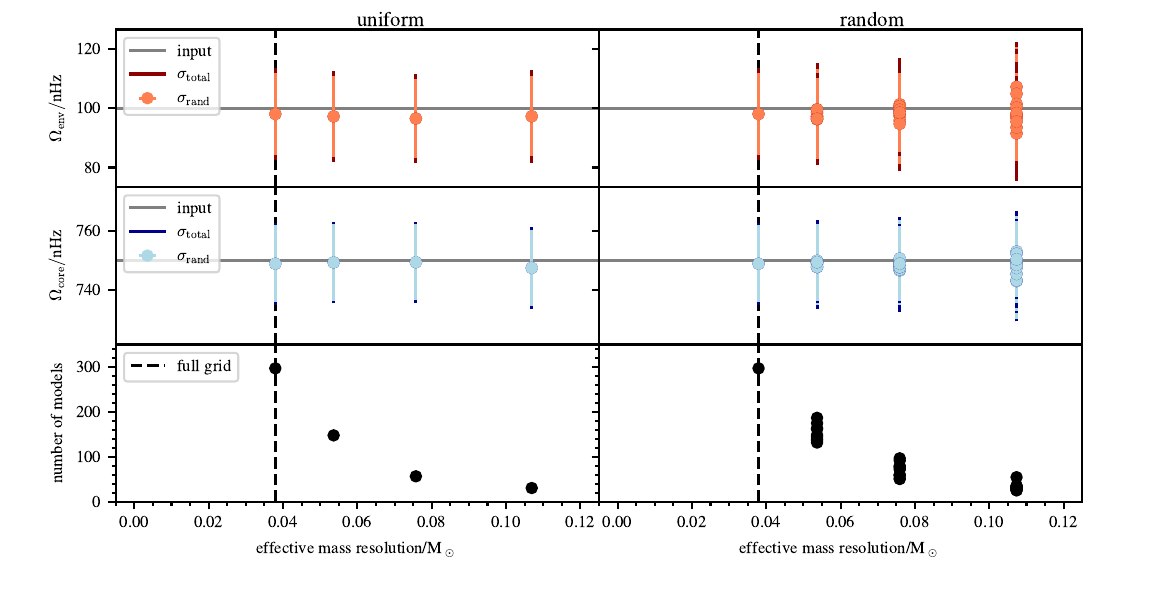}
    \caption{{\it Upper panels:} Estimated envelope rotation as a function of the effective mass resolution. The solid grey line indicates the input value. The vertical black dashed line indicates the mass resolution of the full grid used in the rest of the paper. The left and right column show the results for the uniform and random grid reduction. {\it Middle panels:} Same as the upper panel for the estimated core rotation rates. {\it Lower panel:} Number of models selected for the ensemble inversion as a function of the effective mass resolution.}
    \label{figmassres}
\end{figure*}

The results of the ensemble inversion depend on the properties of the grid from which the reference models get selected. In this subsection, we test how our results depend on the grid density and the physics included in the grid. To test the dependence on the grid density we reduced the number of tracks from the $M, \alpha_{\rm MLT}$-grid by only keeping every $n$-th mass value for $n=2,4,8$. In this way, the originally uniform distribution of $M$ and $\alpha_{\rm MLT}$ values is approximately retained. We summarise the ensemble inversion results for the uniform reduction of the grid in the left column of Fig.~\ref{figmassres} as a function of the effective mass resolution, which we compute as the mass range divided by the square-root of the number of models on the grid. We find that in all cases the input values are recovered within random errors. The random and reference model uncertainties stay approximately the same regardless of the grid density. The number of models selected for the ensemble inversion scales approximately with the number of models on the grid (see lower panel of Fig.~\ref{figmassres}). This conclusion is supported by the random and reference model uncertainties obtained from the much smaller $M$-grid. Hence, the ensemble estimate is not strongly affected by the effective mass resolution in the case of uniform parameter distributions.

We repeated the reduction of the number of reference models by selecting randomly one $n$-th of the models from the $M, \alpha_{\rm MLT}$-grid. The $M$ and $\alpha_{\rm MLT}$ distributions of the latter grids may deviate more strongly from the originally uniform distributions. The results for the random reduction of the grid are shown in the right column of Fig.~\ref{figmassres}. We again recover the input rotation rates in all cases. However, when selecting models from the grids with randomly selected $M$ and $\alpha_{\rm MLT}$ values we find larger deviations from the estimated value for the full grid for decreasing grid resolution. We have generated ten random reduced grids for each reduction factor to illustrate the spread introduced by the non-uniform sampling of the parameter space. Initially, the uniform distribution is achieved by the Sobol sequences and maintained when keeping every $n$-th track. This is not the case when selecting random models from the grid. We conclude that the uniform distribution of $M$ and $\alpha_{\rm MLT}$ values is important to obtain an unbiased estimate of the rotation rates.

The results of stellar evolution calculations are always subject to the assumptions on the physics that were made. On both of our grids we computed the models excluding the effects of rotation. However, the stars under consideration are rotating which may impact the ensemble inversion results. Here, we used MESA to compute a synthetic observation including the effects of rotation. To incorporate the effects of rotation we include the angular momentum transport by viscosity (including contributions from shear, electrons and radiation by default) and increased its efficiency by a factor to obtain rotation rates comparable to observed red-giant stars. For the other parameters, we used the same values as for the fiducial model. We have selected a stellar model from this track with a large frequency separation of $\Delta\nu\approx16~\mu$Hz similar to our fiducial model. We show the internal rotation profile of this model in Fig.~\ref{figrotprofiles}. The synthetic data were computed as described in Sect.~\ref{secsynth}. In order  to roughly reproduce the default core and envelope rotation rates of the envelope step synthetic rotation profile, we increased the efficiency of the angular momentum transport by a factor of 2.16$\cdot10^3$ and set the initial rotation rate on the pre main-sequence to $\Omega_{\rm ini}=11$~nHz. In Fig.~\ref{figrotprofiles} we show that the rotation profile from the MESA model has a fast rotating region that is yet smaller than in the core-step profile. Hence, we would expect an estimate of the core rotation rate from the ensemble inversion that is even more biased to lower rotation rates than for the core-step profile.

We applied the ensemble inversion method to the rotating model in the same way as for all other synthetic observations. Here, we find the following values for the core and envelope rotation rates:
\begin{align*}
\Omega_{\rm core} &= (449 \pm 14_{\rm rand} \pm 3_{\rm ref})~{\rm nHz}\\
\Omega_{\rm env} &= (102 \pm 15_{\rm rand} \pm 4_{\rm ref})~{\rm nHz}\,.
\end{align*}
While the envelope rotation rate is recovered within the uncertainties, we find a significant deviation for the core rotation rate that is larger than for the core-step profile. As we have demonstrated for the core-step profile, this bias towards lower rotation rates is not a consequence of the ensemble inversion, but rather of the shape of the {\it g}-component of the individual rotational kernels that shows substantial sensitivity in the slowly rotating region. As the MESA rotation profile is even more centrally concentrated than the core-step profile, a larger fraction of the sensitivity of the core averaging kernel is sensitive to the slow rotating region, which further decreases the estimate of the core rotation rate. This can be again seen by comparing the transition in the rotation profile and the increase of the core averaging kernel in Fig.~\ref{figrotprofiles}. The rotation profile decreases already well before the core averaging kernel has gained its total sensitivity. To confirm the result of the ensemble inversion we have carried out an inversion using the stellar model used to generate the synthetic observation as a reference model and obtain a nearly identical core rotation rate. We also find a nearly identical result when using MOLA or SOLA inversions instead of eMOLA. Finally, when varying the core-to-envelope contrast of the MESA rotation profile the inversion results behave analogous to the results for the envelope or core step profiles (see Table~\ref{tablerotcomp}). This shows that the selection of reference models does also work robustly on a rotation profile that is physically more realistic, even though the inversion results of the individual reference models in the ensemble suffer from larger errors. We can hence conclude that the inclusion (or the neglect) of rotation in the reference stellar models does not have a significant impact on the ensemble inversion results.
\section{Conclusions}
The accurate measurement of internal rotation rates of evolved stars is important in order to test and improve theoretical models of rotation. Internal rotation rate measurements by means of rotational inversions rely on a stellar reference model that needs to reproduce the structure of the observed star well enough. In this work, we investigate the uncertainties in estimated internal rotation rates of red-giant stars occurring due to structural differences between the observed star and the stellar reference models. We test the dependence of the rotational inversion results on the different stellar parameters by inverting for different sets of synthetic observations. 

Instead of relying on a single reference model that needs to reproduce the rotational kernels of the observations perfectly, we propose the use of a set of reference models that is selected based on global stellar properties ($\Delta\nu$ and $\nu_{\rm max}$), radial mode properties ($\chi^2_{\rm rad}$) and dipole mode properties ($\chi^2_{\rm dip}$ and $\rho$). As described above, it is crucial to match the sensitivity of the observed modes and the modes in the reference model to obtain accurate rotational inversion results. Here, it is especially important that the reference model reproduces the mixed mode character of the observed star. We therefore propose to select reference models for the rotational inversions based on the matching of the mixed mode pattern between the reference model and the observed star. We quantify this match by computing the correlation coefficient $\rho$ between the mode trapping parameter $\zeta$ in the reference model and the observed rotational splittings (Fig.~\ref{figcorrinertia}).

We find that for a splitting correlation coefficient larger than 0.98, the  core and envelope rotation rates are estimated within a narrow region around the input value. We then estimate the internal rotation rates by taking the average of all estimated rotation rates obtained from reference models with $\rho>0.98$ (Fig.~\ref{figensemblesurf}). The standard deviation of the distribution of rotation rates is used as the uncertainty introduced by a discrepancy in structure between the reference models and the observed star. We show that this procedure recovers the input rotation rates within the random uncertainties of the inversion method. We find that the reference models selected span a larger range in masses. This confirms the results of \cite{dimauro2016} of recovering rotation rates across a larger mass range. It shows further that it is not necessary to reproduce all properties of the observed star with a single reference model. Instead, we can rely on a set of reference models that reproduce, on average, the properties of the observed star needed to determine the internal rotation rates.

By using a synthetic observation based on a stellar model with a surface perturbation, we demonstrated that the ensemble inversion result is not strongly affected by this surface perturbation. This can be explained by the fact that the selected reference models implicitly reproduce the sensitivities to the internal rotation of the observed star---a prerequisite to compute accurate rotational inversion results. This property of the ensemble inversion method bears the potential to estimate internal rotation rates of red-giant stars without the need to correct for the surface effect on a star by star basis.

To test the dependence on different stellar parameters, we constructed different sets of synthetic observations with varying the stellar mass, the mixing length parameter, the chemical composition and the age of the star. The results are summarised in Fig.~\ref{figsummary}. We note that across all synthetic observations, the envelope rotation rate tends to be overestimated while the core rotation rate tends to be underestimated. This occurs because a mismatch of the sensitivity between observed star and reference models will more likely match a mode with more core sensitivity to a mode with less core sensitivity and vice versa. In the inversion process, a {\it p}-dominated splitting might therefore be interpreted as a {\it g}-dominated one and therefore decrease the estimate of the core rotation rate, or a {\it g}-dominated mode might be interpreted as a {\it p}-dominated mode and therefore increase the estimate of the envelope rotation rate. We find that the chemical composition needs to be constrained to within $\pm0.1$~dex to recover accurate core and envelope rotation rates.

Figure~\ref{figalphamassfinal} shows that most of the models in the selected set of reference models do not reproduce the observed star. Nevertheless, they are able to reproduce the input rotation rates. This can be explained by the fact that the sensitivity of the dipole mode rotational kernels is largely dominated by two components \citep[][Sect. 6.2]{ahlborn2022}---a core and an envelope component. Among the individual rotational kernels of a stellar model, it is only the weight of the core to the envelope part that is varying. As long as a model is reproducing the core to envelope weights of the rotational kernels of the observed star, the rotation rates can be recovered accurately---disregarding differences in other fundamental stellar parameters.

\begin{acknowledgements}
We would like to thank the anonymous referee for helpful comments to improve this manuscript. FA would like to thank Joel Ong for insightful discussions that helped improving the manuscript.  The research leading to the presented results has received funding from the European Research Council under the European Community’s Horizon 2020 Framework/ERC grant agreement no 101000296 (DipolarSounds). Funding for the Stellar Astrophysics Centre is provided by The Danish National Research Foundation (Grant agreement no.: DNRF106). SB acknowledges NSF grant AST-2205026. She would like to thank the Heidelberg Institute of Theoretical Studies for their hospitality at the time that this project was conceived.
\end{acknowledgements}

%
   \bibliographystyle{aa} 
   \bibliography{bibliography.bib} 

\begin{thebibliography}{79}
\expandafter\ifx\csname natexlab\endcsname\relax\def\natexlab#1{#1}\fi

\bibitem[{{Aerts} {et~al.}(2010){Aerts}, {Christensen-Dalsgaard}, \&
  {Kurtz}}]{aerts2010}
{Aerts}, C., {Christensen-Dalsgaard}, J., \& {Kurtz}, D.~W. 2010,
  {Asteroseismology}

\bibitem[{{Aerts} {et~al.}(2019){Aerts}, {Mathis}, \& {Rogers}}]{aerts2019}
{Aerts}, C., {Mathis}, S., \& {Rogers}, T.~M. 2019, \araa, 57, 35

\bibitem[{{Ahlborn} {et~al.}(2020){Ahlborn}, {Bellinger}, {Hekker}, {Basu}, \&
  {Angelou}}]{ahlborn2020}
{Ahlborn}, F., {Bellinger}, E.~P., {Hekker}, S., {Basu}, S., \& {Angelou},
  G.~C. 2020, \aap, 639, A98

\bibitem[{{Ahlborn} {et~al.}(2022){Ahlborn}, {Bellinger}, {Hekker}, {Basu}, \&
  {Mokrytska}}]{ahlborn2022}
{Ahlborn}, F., {Bellinger}, E.~P., {Hekker}, S., {Basu}, S., \& {Mokrytska}, D.
  2022, \aap, 668, A98

\bibitem[{{Alvan} {et~al.}(2013){Alvan}, {Mathis}, \& {Decressin}}]{alvan2013}
{Alvan}, L., {Mathis}, S., \& {Decressin}, T. 2013, \aap, 553, A86

\bibitem[{{Asplund} {et~al.}(2009){Asplund}, {Grevesse}, {Sauval}, \&
  {Scott}}]{asplund2009}
{Asplund}, M., {Grevesse}, N., {Sauval}, A.~J., \& {Scott}, P. 2009, \araa, 47,
  481

\bibitem[{{Backus} \& {Gilbert}(1968)}]{backus1968}
{Backus}, G. \& {Gilbert}, F. 1968, Geophysical Journal, 16, 169

\bibitem[{{Ball} \& {Gizon}(2014)}]{ball2014}
{Ball}, W.~H. \& {Gizon}, L. 2014, \aap, 568, A123

\bibitem[{{Ball} {et~al.}(2018){Ball}, {Theme{\ss}l}, \& {Hekker}}]{ball2018}
{Ball}, W.~H., {Theme{\ss}l}, N., \& {Hekker}, S. 2018, \mnras, 478, 4697

\bibitem[{{Beck} {et~al.}(2011){Beck}, {Bedding}, {Mosser}, {Stello}, {Garcia},
  {Kallinger}, {Hekker}, {Elsworth}, {Frandsen}, {Carrier}, {De Ridder},
  {Aerts}, {White}, {Huber}, {Dupret}, {Montalb{\'a}n}, {Miglio}, {Noels},
  {Chaplin}, {Kjeldsen}, {Christensen-Dalsgaard}, {Gilliland}, {Brown},
  {Kawaler}, {Mathur}, \& {Jenkins}}]{beck2011}
{Beck}, P.~G., {Bedding}, T.~R., {Mosser}, B., {et~al.} 2011, Science, 332, 205

\bibitem[{{Beck} {et~al.}(2014){Beck}, {Hambleton}, {Vos}, {Kallinger},
  {Bloemen}, {Tkachenko}, {Garc{\'{\i}}a}, {{\O}stensen}, {Aerts}, {Kurtz}, {De
  Ridder}, {Hekker}, {Pavlovski}, {Mathur}, {De Smedt}, {Derekas}, {Corsaro},
  {Mosser}, {Van Winckel}, {Huber}, {Degroote}, {Davies}, {Pr{\v s}a},
  {Debosscher}, {Elsworth}, {Nemeth}, {Siess}, {Schmid}, {P{\'a}pics}, {de
  Vries}, {van Marle}, {Marcos-Arenal}, \& {Lobel}}]{beck2014}
{Beck}, P.~G., {Hambleton}, K., {Vos}, J., {et~al.} 2014, \aap, 564, A36

\bibitem[{{Beck} {et~al.}(2018){Beck}, {Kallinger}, {Pavlovski}, {Palacios},
  {Tkachenko}, {Mathis}, {Garc{\'\i}a}, {Corsaro}, {Johnston}, {Mosser},
  {Ceillier}, {do Nascimento}, \& {Raskin}}]{beck2018}
{Beck}, P.~G., {Kallinger}, T., {Pavlovski}, K., {et~al.} 2018, \aap, 612, A22

\bibitem[{{Beck} {et~al.}(2012){Beck}, {Montalban}, {Kallinger}, {De Ridder},
  {Aerts}, {Garc{\'{\i}}a}, {Hekker}, {Dupret}, {Mosser}, {Eggenberger},
  {Stello}, {Elsworth}, {Frandsen}, {Carrier}, {Hillen}, {Gruberbauer},
  {Christensen-Dalsgaard}, {Miglio}, {Valentini}, {Bedding}, {Kjeldsen},
  {Girouard}, {Hall}, \& {Ibrahim}}]{beck2012}
{Beck}, P.~G., {Montalban}, J., {Kallinger}, T., {et~al.} 2012, Nature, 481, 55

\bibitem[{{Bedding} {et~al.}(2011){Bedding}, {Mosser}, {Huber},
  {Montalb{\'a}n}, {Beck}, {Christensen-Dalsgaard}, {Elsworth},
  {Garc{\'{\i}}a}, {Miglio}, {Stello}, {White}, {De Ridder}, {Hekker}, {Aerts},
  {Barban}, {Belkacem}, {Broomhall}, {Brown}, {Buzasi}, {Carrier}, {Chaplin},
  {di Mauro}, {Dupret}, {Frandsen}, {Gilliland}, {Goupil}, {Jenkins},
  {Kallinger}, {Kawaler}, {Kjeldsen}, {Mathur}, {Noels}, {Silva Aguirre}, \&
  {Ventura}}]{bedding2011}
{Bedding}, T.~R., {Mosser}, B., {Huber}, D., {et~al.} 2011, \nat, 471, 608

\bibitem[{{Belkacem} {et~al.}(2015{\natexlab{a}}){Belkacem}, {Marques},
  {Goupil}, {Mosser}, {Sonoi}, {Ouazzani}, {Dupret}, {Mathis}, \&
  {Grosjean}}]{belkacem2015b}
{Belkacem}, K., {Marques}, J.~P., {Goupil}, M.~J., {et~al.} 2015{\natexlab{a}},
  \aap, 579, A31

\bibitem[{{Belkacem} {et~al.}(2015{\natexlab{b}}){Belkacem}, {Marques},
  {Goupil}, {Sonoi}, {Ouazzani}, {Dupret}, {Mathis}, {Mosser}, \&
  {Grosjean}}]{belkacem2015a}
{Belkacem}, K., {Marques}, J.~P., {Goupil}, M.~J., {et~al.} 2015{\natexlab{b}},
  \aap, 579, A30

\bibitem[{{Bellinger} {et~al.}(2016){Bellinger}, {Angelou}, {Hekker}, {Basu},
  {Ball}, \& {Guggenberger}}]{bellinger2016}
{Bellinger}, E.~P., {Angelou}, G.~C., {Hekker}, S., {et~al.} 2016, \apj, 830,
  31

\bibitem[{{B{\"o}hm-Vitense}(1958)}]{boehm1958}
{B{\"o}hm-Vitense}, E. 1958, \zap, 46, 108

\bibitem[{{Bouchy} \& {Carrier}(2001)}]{bouchy2001}
{Bouchy}, F. \& {Carrier}, F. 2001, \aap, 374, L5

\bibitem[{{Brown}(1984)}]{brown1984}
{Brown}, T.~M. 1984, Science, 226, 687

\bibitem[{{Buldgen} {et~al.}(2024){Buldgen}, {Fellay}, {B{\'e}trisey},
  {Deheuvels}, {Farnir}, \& {Farrell}}]{buldgen2024}
{Buldgen}, G., {Fellay}, L., {B{\'e}trisey}, J., {et~al.} 2024, \aap, 689, A307

\bibitem[{{Cantiello} {et~al.}(2014){Cantiello}, {Mankovich}, {Bildsten},
  {Christensen-Dalsgaard}, \& {Paxton}}]{cantiello2014}
{Cantiello}, M., {Mankovich}, C., {Bildsten}, L., {Christensen-Dalsgaard}, J.,
  \& {Paxton}, B. 2014, The Astrophysical Journal, 788, 93

\bibitem[{{Ceillier} {et~al.}(2013){Ceillier}, {Eggenberger}, {Garc{\'{\i}}a},
  \& {Mathis}}]{ceillier2013}
{Ceillier}, T., {Eggenberger}, P., {Garc{\'{\i}}a}, R.~A., \& {Mathis}, S.
  2013, \aap, 555, A54

\bibitem[{{Choi} {et~al.}(2016){Choi}, {Dotter}, {Conroy}, {Cantiello},
  {Paxton}, \& {Johnson}}]{choi2016}
{Choi}, J., {Dotter}, A., {Conroy}, C., {et~al.} 2016, \apj, 823, 102

\bibitem[{{Christensen-Dalsgaard} {et~al.}(1988){Christensen-Dalsgaard},
  {Dappen}, \& {Lebreton}}]{christensen1988}
{Christensen-Dalsgaard}, J., {Dappen}, W., \& {Lebreton}, Y. 1988, \nat, 336,
  634

\bibitem[{{Christensen-Dalsgaard} {et~al.}(1990){Christensen-Dalsgaard},
  {Schou}, \& {Thompson}}]{christensen1990}
{Christensen-Dalsgaard}, J., {Schou}, J., \& {Thompson}, M.~J. 1990, Monthly
  Notices of the Royal Astronomical Society, 242, 353

\bibitem[{{Deheuvels} {et~al.}(2015){Deheuvels}, {Ballot}, {Beck}, {Mosser},
  {{\O}stensen}, {Garc{\'{\i}}a}, \& {Goupil}}]{deheuvels2015}
{Deheuvels}, S., {Ballot}, J., {Beck}, P.~G., {et~al.} 2015, Astronomy and
  Astrophysics, 580, A96

\bibitem[{{Deheuvels} {et~al.}(2020){Deheuvels}, {Ballot}, {Eggenberger},
  {Spada}, {Noll}, \& {den Hartogh}}]{deheuvels2020}
{Deheuvels}, S., {Ballot}, J., {Eggenberger}, P., {et~al.} 2020, \aap, 641,
  A117

\bibitem[{{Deheuvels} {et~al.}(2014){Deheuvels}, {Do{\u g}an}, {Goupil},
  {Appourchaux}, {Benomar}, {Bruntt}, {Campante}, {Casagrande}, {Ceillier},
  {Davies}, {De Cat}, {Fu}, {Garc{\'{\i}}a}, {Lobel}, {Mosser}, {Reese},
  {Regulo}, {Schou}, {Stahn}, {Thygesen}, {Yang}, {Chaplin},
  {Christensen-Dalsgaard}, {Eggenberger}, {Gizon}, {Mathis},
  {Molenda-{\.Z}akowicz}, \& {Pinsonneault}}]{deheuvels2014}
{Deheuvels}, S., {Do{\u g}an}, G., {Goupil}, M.~J., {et~al.} 2014, Astronomy
  and Astrophysics, 564, A27

\bibitem[{{Deheuvels} {et~al.}(2012){Deheuvels}, {Garc{\'{\i}}a}, {Chaplin},
  {Basu}, {Antia}, {Appourchaux}, {Benomar}, {Davies}, {Elsworth}, {Gizon},
  {Goupil}, {Reese}, {Regulo}, {Schou}, {Stahn}, {Casagrande},
  {Christensen-Dalsgaard}, {Fischer}, {Hekker}, {Kjeldsen}, {Mathur}, {Mosser},
  {Pinsonneault}, {Valenti}, {Christiansen}, {Kinemuchi}, \&
  {Mullally}}]{deheuvels2012}
{Deheuvels}, S., {Garc{\'{\i}}a}, R.~A., {Chaplin}, W.~J., {et~al.} 2012, The
  Astrophysical Journal, 756, 19

\bibitem[{{Deheuvels} {et~al.}(2017){Deheuvels}, {Ouazzani}, \&
  {Basu}}]{deheuvels2017}
{Deheuvels}, S., {Ouazzani}, R.~M., \& {Basu}, S. 2017, \aap, 605, A75

\bibitem[{{Di Mauro} {et~al.}(2016){Di Mauro}, {Ventura}, {Cardini}, {Stello},
  {Christensen-Dalsgaard}, {Dziembowski}, {Patern{\`o}}, {Beck}, {Bloemen},
  {Davies}, {De Smedt}, {Elsworth}, {Garc{\'{\i}}a}, {Hekker}, {Mosser}, \&
  {Tkachenko}}]{dimauro2016}
{Di Mauro}, M.~P., {Ventura}, R., {Cardini}, D., {et~al.} 2016, The
  Astrophysical Journal, 817, 65

\bibitem[{{Di Mauro} {et~al.}(2018){Di Mauro}, {Ventura}, {Corsaro}, \&
  {Lustosa De Moura}}]{dimauro2018}
{Di Mauro}, M.~P., {Ventura}, R., {Corsaro}, E., \& {Lustosa De Moura}, B.
  2018, \apj, 862, 9

\bibitem[{{Dupret} {et~al.}(2009){Dupret}, {Belkacem}, {Samadi}, {Montalban},
  {Moreira}, {Miglio}, {Godart}, {Ventura}, {Ludwig}, {Grigahc{\`e}ne},
  {Goupil}, {Noels}, \& {Caffau}}]{dupret2009}
{Dupret}, M.~A., {Belkacem}, K., {Samadi}, R., {et~al.} 2009, \aap, 506, 57

\bibitem[{{Eddington}(1926)}]{eddington1926}
{Eddington}, A.~S. 1926, {The Internal Constitution of the Stars} (Cambridge
  University Press)

\bibitem[{{Eggenberger} {et~al.}(2019){Eggenberger}, {den Hartogh}, {Buldgen},
  {Meynet}, {Salmon}, \& {Deheuvels}}]{eggenberger2019}
{Eggenberger}, P., {den Hartogh}, J.~W., {Buldgen}, G., {et~al.} 2019, \aap,
  631, L6

\bibitem[{{Eggenberger} {et~al.}(2010){Eggenberger}, {Miglio}, {Montalban},
  {Moreira}, {Noels}, {Meynet}, \& {Maeder}}]{eggenberger2010}
{Eggenberger}, P., {Miglio}, A., {Montalban}, J., {et~al.} 2010, Astronomy and
  Astrophysics, 509, A72

\bibitem[{{Eggenberger} {et~al.}(2012){Eggenberger}, {Montalb{\'a}n}, \&
  {Miglio}}]{eggenberger2012}
{Eggenberger}, P., {Montalb{\'a}n}, J., \& {Miglio}, A. 2012, The Astrophysical
  Journal, 544, L4

\bibitem[{Fellay {et~al.}(2021)Fellay, Buldgen, Eggenberger, Khan, Salmon,
  Miglio, \& Montalbán}]{fellay2021}
Fellay, L., Buldgen, G., Eggenberger, P., {et~al.} 2021, 654, A133, {ADS}
  Bibcode: 2021A\&A...654A.133F

\bibitem[{{Ferguson} {et~al.}(2005){Ferguson}, {Alexander}, {Allard}, {Barman},
  {Bodnarik}, {Hauschildt}, {Heffner-Wong}, \& {Tamanai}}]{ferguson2005}
{Ferguson}, J.~W., {Alexander}, D.~R., {Allard}, F., {et~al.} 2005, \apj, 623,
  585

\bibitem[{{Fuller} {et~al.}(2014){Fuller}, {Lecoanet}, {Cantiello}, \&
  {Brown}}]{fuller2014}
{Fuller}, J., {Lecoanet}, D., {Cantiello}, M., \& {Brown}, B. 2014, \apj, 796,
  17

\bibitem[{{Fuller} {et~al.}(2019){Fuller}, {Piro}, \& {Jermyn}}]{fuller2019}
{Fuller}, J., {Piro}, A.~L., \& {Jermyn}, A.~S. 2019, \mnras, 485, 3661

\bibitem[{{Gehan} {et~al.}(2018){Gehan}, {Mosser}, {Michel}, {Samadi}, \&
  {Kallinger}}]{gehan2018}
{Gehan}, C., {Mosser}, B., {Michel}, E., {Samadi}, R., \& {Kallinger}, T. 2018,
  \aap, 616, A24

\bibitem[{{Gizon} \& {Solanki}(2003)}]{gizon2003}
{Gizon}, L. \& {Solanki}, S.~K. 2003, The Astrophysical Journal, 589, 1009

\bibitem[{{Gough}(1985)}]{gough1985}
{Gough}, D. 1985, \solphys, 100, 65

\bibitem[{{Goupil} {et~al.}(2013){Goupil}, {Mosser}, {Marques}, {Ouazzani},
  {Belkacem}, {Lebreton}, \& {Samadi}}]{goupil2013}
{Goupil}, M.~J., {Mosser}, B., {Marques}, J.~P., {et~al.} 2013, Astronomy and
  Astrophysics, 549, A75

\bibitem[{{Grevesse} \& {Sauval}(1998)}]{grevesse1998}
{Grevesse}, N. \& {Sauval}, A.~J. 1998, \ssr, 85, 161

\bibitem[{{Houdek} {et~al.}(2017){Houdek}, {Trampedach}, {Aarslev}, \&
  {Christensen-Dalsgaard}}]{houdek2017}
{Houdek}, G., {Trampedach}, R., {Aarslev}, M.~J., \& {Christensen-Dalsgaard},
  J. 2017, \mnras, 464, L124

\bibitem[{{Iglesias} \& {Rogers}(1996)}]{iglesias1996}
{Iglesias}, C.~A. \& {Rogers}, F.~J. 1996, \apj, 464, 943

\bibitem[{{Jermyn} {et~al.}(2023){Jermyn}, {Bauer}, {Schwab}, {Farmer}, {Ball},
  {Bellinger}, {Dotter}, {Joyce}, {Marchant}, {Mombarg}, {Wolf}, {Sunny Wong},
  {Cinquegrana}, {Farrell}, {Smolec}, {Thoul}, {Cantiello}, {Herwig}, {Toloza},
  {Bildsten}, {Townsend}, \& {Timmes}}]{jermyn2023}
{Jermyn}, A.~S., {Bauer}, E.~B., {Schwab}, J., {et~al.} 2023, \apjs, 265, 15

\bibitem[{{J{\o}rgensen} {et~al.}(2018){J{\o}rgensen}, {Mosumgaard}, {Weiss},
  {Silva Aguirre}, \& {Christensen-Dalsgaard}}]{jorgensen2018}
{J{\o}rgensen}, A. C.~S., {Mosumgaard}, J.~R., {Weiss}, A., {Silva Aguirre},
  V., \& {Christensen-Dalsgaard}, J. 2018, \mnras, 481, L35

\bibitem[{{Kjeldsen} \& {Bedding}(1995)}]{kjeldsen1995}
{Kjeldsen}, H. \& {Bedding}, T.~R. 1995, Astronomy and Astrophysics, 293, 87

\bibitem[{{Kjeldsen} {et~al.}(2008){Kjeldsen}, {Bedding}, \&
  {Christensen-Dalsgaard}}]{kjeldsen2008}
{Kjeldsen}, H., {Bedding}, T.~R., \& {Christensen-Dalsgaard}, J. 2008, \apjl,
  683, L175

\bibitem[{{Klion} \& {Quataert}(2017)}]{klion2017}
{Klion}, H. \& {Quataert}, E. 2017, \mnras, 464, L16

\bibitem[{Kuhn(1955)}]{kuhn1955}
Kuhn, H.~W. 1955, Naval Research Logistics Quarterly, 2, 83

\bibitem[{{Maeder}(2009)}]{maeder2009}
{Maeder}, A. 2009, {Physics, Formation and Evolution of Rotating Stars}
  (Springer)

\bibitem[{{Marques} {et~al.}(2013){Marques}, {Goupil}, {Lebreton}, {Talon},
  {Palacios}, {Belkacem}, {Ouazzani}, {Mosser}, {Moya}, {Morel}, {Pichon},
  {Mathis}, {Zahn}, {Turck-Chi{\`e}ze}, \& {Nghiem}}]{marques2013}
{Marques}, J.~P., {Goupil}, M.~J., {Lebreton}, Y., {et~al.} 2013, \aap, 549,
  A74

\bibitem[{{Mosser} {et~al.}(2024){Mosser}, {Dr{\'e}au}, {Pin{\c{c}}on},
  {Deheuvels}, {Belkacem}, {Lebreton}, {Goupil}, \& {Michel}}]{mosser2024}
{Mosser}, B., {Dr{\'e}au}, G., {Pin{\c{c}}on}, C., {et~al.} 2024, \aap, 681,
  L20

\bibitem[{{Mosser} {et~al.}(2012{\natexlab{a}}){Mosser}, {Elsworth}, {Hekker},
  {Huber}, {Kallinger}, {Mathur}, {Belkacem}, {Goupil}, {Samadi}, {Barban},
  {Bedding}, {Chaplin}, {Garc{\'\i}a}, {Stello}, {De Ridder}, {Middour},
  {Morris}, \& {Quintana}}]{mosser2012c}
{Mosser}, B., {Elsworth}, Y., {Hekker}, S., {et~al.} 2012{\natexlab{a}}, \aap,
  537, A30

\bibitem[{{Mosser} {et~al.}(2012{\natexlab{b}}){Mosser}, {Goupil}, {Belkacem},
  {Marques}, {Beck}, {Bloemen}, {De Ridder}, {Barban}, {Deheuvels}, {Elsworth},
  {Hekker}, {Kallinger}, {Ouazzani}, {Pinsonneault}, {Samadi}, {Stello},
  {Garc{\'{\i}}a}, {Klaus}, {Li}, {Mathur}, \& {Morris}}]{mosser2012}
{Mosser}, B., {Goupil}, M.~J., {Belkacem}, K., {et~al.} 2012{\natexlab{b}},
  Astronomy and Astrophysics, 548, A10

\bibitem[{{Ong}(2024)}]{ong2024}
{Ong}, J.~M.~J. 2024, \apj, 960, 2

\bibitem[{{Ong} {et~al.}(2021){Ong}, {Basu}, \& {Roxburgh}}]{ong2021}
{Ong}, J.~M.~J., {Basu}, S., \& {Roxburgh}, I.~W. 2021, \apj, 920, 8

\bibitem[{{Paxton} {et~al.}(2011){Paxton}, {Bildsten}, {Dotter}, {Herwig},
  {Lesaffre}, \& {Timmes}}]{paxton2011}
{Paxton}, B., {Bildsten}, L., {Dotter}, A., {et~al.} 2011, \apjs, 192, 3,
  available at \url{http://mesa.sourceforge.net/}

\bibitem[{{Paxton} {et~al.}(2013){Paxton}, {Cantiello}, {Arras}, {Bildsten},
  {Brown}, {Dotter}, {Mankovich}, {Montgomery}, {Stello}, {Timmes}, \&
  {Townsend}}]{paxton2013}
{Paxton}, B., {Cantiello}, M., {Arras}, P., {et~al.} 2013, \apjs, 208, 4

\bibitem[{{Paxton} {et~al.}(2015){Paxton}, {Marchant}, {Schwab}, {Bauer},
  {Bildsten}, {Cantiello}, {Dessart}, {Farmer}, {Hu}, {Langer}, {Townsend},
  {Townsley}, \& {Timmes}}]{paxton2015}
{Paxton}, B., {Marchant}, P., {Schwab}, J., {et~al.} 2015, \apjs, 220, 15

\bibitem[{{Paxton} {et~al.}(2018){Paxton}, {Schwab}, {Bauer}, {Bildsten},
  {Blinnikov}, {Duffell}, {Farmer}, {Goldberg}, {Marchant}, {Sorokina},
  {Thoul}, {Townsend}, \& {Timmes}}]{paxton2018}
{Paxton}, B., {Schwab}, J., {Bauer}, E.~B., {et~al.} 2018, \apjs, 234, 34

\bibitem[{{Paxton} {et~al.}(2019){Paxton}, {Smolec}, {Schwab}, {Gautschy},
  {Bildsten}, {Cantiello}, {Dotter}, {Farmer}, {Goldberg}, {Jermyn}, {Kanbur},
  {Marchant}, {Thoul}, {Townsend}, {Wolf}, {Zhang}, \& {Timmes}}]{paxton2019}
{Paxton}, B., {Smolec}, R., {Schwab}, J., {et~al.} 2019, The Astrophysical
  Journal Supplement Series, 243, 10

\bibitem[{{Pijpers} \& {Thompson}(1992)}]{pijpers1992}
{Pijpers}, F.~P. \& {Thompson}, M.~J. 1992, \aap, 262, L33

\bibitem[{{Pijpers} \& {Thompson}(1994)}]{pijpers1994}
{Pijpers}, F.~P. \& {Thompson}, M.~J. 1994, \aap, 281, 231

\bibitem[{{Pin{\c{c}}on} {et~al.}(2017){Pin{\c{c}}on}, {Belkacem}, {Goupil}, \&
  {Marques}}]{pincon2017}
{Pin{\c{c}}on}, C., {Belkacem}, K., {Goupil}, M.~J., \& {Marques}, J.~P. 2017,
  \aap, 605, A31

\bibitem[{{Planck Collaboration} {et~al.}(2016){Planck Collaboration}, {Ade, P.
  A. R.}, {Aghanim, N.}, {Arnaud, M.}, {Ashdown, M.}, {Aumont, J.},
  {Baccigalupi, C.}, {Banday, A. J.}, {Barreiro, R. B.}, {Bartlett, J. G.},
  {Bartolo, N.}, {Battaner, E.}, {Battye, R.}, {Benabed, K.}, {Beno\^{\i}t,
  A.}, {Benoit-L\'evy, A.}, {Bernard, J.-P.}, {Bersanelli, M.}, {Bielewicz,
  P.}, {Bock, J. J.}, {Bonaldi, A.}, {Bonavera, L.}, {Bond, J. R.}, {Borrill,
  J.}, {Bouchet, F. R.}, {Boulanger, F.}, {Bucher, M.}, {Burigana, C.},
  {Butler, R. C.}, {Calabrese, E.}, {Cardoso, J.-F.}, {Catalano, A.},
  {Challinor, A.}, {Chamballu, A.}, {Chary, R.-R.}, {Chiang, H. C.}, {Chluba,
  J.}, {Christensen, P. R.}, {Church, S.}, {Clements, D. L.}, {Colombi, S.},
  {Colombo, L. P. L.}, {Combet, C.}, {Coulais, A.}, {Crill, B. P.}, {Curto,
  A.}, {Cuttaia, F.}, {Danese, L.}, {Davies, R. D.}, {Davis, R. J.}, {de
  Bernardis, P.}, {de Rosa, A.}, {de Zotti, G.}, {Delabrouille, J.}, {D\'esert,
  F.-X.}, {Di Valentino, E.}, {Dickinson, C.}, {Diego, J. M.}, {Dolag, K.},
  {Dole, H.}, {Donzelli, S.}, {Dor\'e, O.}, {Douspis, M.}, {Ducout, A.},
  {Dunkley, J.}, {Dupac, X.}, {Efstathiou, G.}, {Elsner, F.}, {En\ss{}lin, T.
  A.}, {Eriksen, H. K.}, {Farhang, M.}, {Fergusson, J.}, {Finelli, F.}, {Forni,
  O.}, {Frailis, M.}, {Fraisse, A. A.}, {Franceschi, E.}, {Frejsel, A.},
  {Galeotta, S.}, {Galli, S.}, {Ganga, K.}, {Gauthier, C.}, {Gerbino, M.},
  {Ghosh, T.}, {Giard, M.}, {Giraud-H\'eraud, Y.}, {Giusarma, E.}, {Gjerl\o{}w,
  E.}, {Gonz\'alez-Nuevo, J.}, {G\'orski, K. M.}, {Gratton, S.}, {Gregorio,
  A.}, {Gruppuso, A.}, {Gudmundsson, J. E.}, {Hamann, J.}, {Hansen, F. K.},
  {Hanson, D.}, {Harrison, D. L.}, {Helou, G.}, {Henrot-Versill\'e, S.},
  {Hern\'andez-Monteagudo, C.}, {Herranz, D.}, {Hildebrandt, S. R.}, {Hivon,
  E.}, {Hobson, M.}, {Holmes, W. A.}, {Hornstrup, A.}, {Hovest, W.}, {Huang,
  Z.}, {Huffenberger, K. M.}, {Hurier, G.}, {Jaffe, A. H.}, {Jaffe, T. R.},
  {Jones, W. C.}, {Juvela, M.}, {Keih\"anen, E.}, {Keskitalo, R.}, {Kisner, T.
  S.}, {Kneissl, R.}, {Knoche, J.}, {Knox, L.}, {Kunz, M.}, {Kurki-Suonio, H.},
  {Lagache, G.}, {L\"ahteenm\"aki, A.}, {Lamarre, J.-M.}, {Lasenby, A.},
  {Lattanzi, M.}, {Lawrence, C. R.}, {Leahy, J. P.}, {Leonardi, R.},
  {Lesgourgues, J.}, {Levrier, F.}, {Lewis, A.}, {Liguori, M.}, {Lilje, P. B.},
  {Linden-V\o{}rnle, M.}, {L\'opez-Caniego, M.}, {Lubin, P. M.},
  {Mac\'{\i}as-P\'erez, J. F.}, {Maggio, G.}, {Maino, D.}, {Mandolesi, N.},
  {Mangilli, A.}, {Marchini, A.}, {Maris, M.}, {Martin, P. G.}, {Martinelli,
  M.}, {Mart\'{\i}nez-Gonz\'alez, E.}, {Masi, S.}, {Matarrese, S.}, {McGehee,
  P.}, {Meinhold, P. R.}, {Melchiorri, A.}, {Melin, J.-B.}, {Mendes, L.},
  {Mennella, A.}, {Migliaccio, M.}, {Millea, M.}, {Mitra, S.},
  {Miville-Desch\^enes, M.-A.}, {Moneti, A.}, {Montier, L.}, {Morgante, G.},
  {Mortlock, D.}, {Moss, A.}, {Munshi, D.}, {Murphy, J. A.}, {Naselsky, P.},
  {Nati, F.}, {Natoli, P.}, {Netterfield, C. B.}, {N\o{}rgaard-Nielsen, H. U.},
  {Noviello, F.}, {Novikov, D.}, {Novikov, I.}, {Oxborrow, C. A.}, {Paci, F.},
  {Pagano, L.}, {Pajot, F.}, {Paladini, R.}, {Paoletti, D.}, {Partridge, B.},
  {Pasian, F.}, {Patanchon, G.}, {Pearson, T. J.}, {Perdereau, O.}, {Perotto,
  L.}, {Perrotta, F.}, {Pettorino, V.}, {Piacentini, F.}, {Piat, M.},
  {Pierpaoli, E.}, {Pietrobon, D.}, {Plaszczynski, S.}, {Pointecouteau, E.},
  {Polenta, G.}, {Popa, L.}, {Pratt, G. W.}, {Pr\'ezeau, G.}, {Prunet, S.},
  {Puget, J.-L.}, {Rachen, J. P.}, {Reach, W. T.}, {Rebolo, R.}, {Reinecke,
  M.}, {Remazeilles, M.}, {Renault, C.}, {Renzi, A.}, {Ristorcelli, I.},
  {Rocha, G.}, {Rosset, C.}, {Rossetti, M.}, {Roudier, G.}, {Rouill\'e
  d\'{}Orfeuil, B.}, {Rowan-Robinson, M.}, {Rubi\~no-Mart\'{\i}n, J. A.},
  {Rusholme, B.}, {Said, N.}, {Salvatelli, V.}, {Salvati, L.}, {Sandri, M.},
  {Santos, D.}, {Savelainen, M.}, {Savini, G.}, {Scott, D.}, {Seiffert, M. D.},
  {Serra, P.}, {Shellard, E. P. S.}, {Spencer, L. D.}, {Spinelli, M.},
  {Stolyarov, V.}, {Stompor, R.}, {Sudiwala, R.}, {Sunyaev, R.}, {Sutton, D.},
  {Suur-Uski, A.-S.}, {Sygnet, J.-F.}, {Tauber, J. A.}, {Terenzi, L.},
  {Toffolatti, L.}, {Tomasi, M.}, {Tristram, M.}, {Trombetti, T.}, {Tucci, M.},
  {Tuovinen, J.}, {T\"urler, M.}, {Umana, G.}, {Valenziano, L.}, {Valiviita,
  J.}, {Van Tent, F.}, {Vielva, P.}, {Villa, F.}, {Wade, L. A.}, {Wandelt, B.
  D.}, {Wehus, I. K.}, {White, M.}, {White, S. D. M.}, {Wilkinson, A.}, {Yvon,
  D.}, {Zacchei, A.}, \& {Zonca, A.}}]{planck}
{Planck Collaboration}, {Ade, P. A. R.}, {Aghanim, N.}, {et~al.} 2016, A\&A,
  594, A13

\bibitem[{{Schou} {et~al.}(1998){Schou}, {Antia}, {Basu}, {Bogart}, {Bush},
  {Chitre}, {Christensen-Dalsgaard}, {Di Mauro}, {Dziembowski}, {Eff-Darwich},
  {Gough}, {Haber}, {Hoeksema}, {Howe}, {Korzennik}, {Kosovichev}, {Larsen},
  {Pijpers}, {Scherrer}, {Sekii}, {Tarbell}, {Title}, {Thompson}, \&
  {Toomre}}]{schou1998}
{Schou}, J., {Antia}, H.~M., {Basu}, S., {et~al.} 1998, \apj, 505, 390

\bibitem[{{Schunker} {et~al.}(2016){Schunker}, {Schou}, \&
  {Ball}}]{schunker2016a}
{Schunker}, H., {Schou}, J., \& {Ball}, W.~H. 2016, \aap, 586, A24

\bibitem[{{Sobol'}(1967)}]{sobol1967}
{Sobol'}, I.~M. 1967, USSR Computational Mathematics and Mathematical Physics,
  7, 86

\bibitem[{{Spruit}(2002)}]{spruit2002}
{Spruit}, H.~C. 2002, \aap, 381, 923

\bibitem[{{Theme{\ss}l} {et~al.}(2018){Theme{\ss}l}, {Hekker}, {Southworth},
  {Beck}, {Pavlovski}, {Tkachenko}, {Angelou}, {Ball}, {Barban}, {Corsaro},
  {Elsworth}, {Handberg}, \& {Kallinger}}]{themessl2018}
{Theme{\ss}l}, N., {Hekker}, S., {Southworth}, J., {et~al.} 2018, \mnras, 478,
  4669

\bibitem[{{Townsend} {et~al.}(2018){Townsend}, {Goldstein}, \&
  {Zweibel}}]{townsend2018}
{Townsend}, R.~H.~D., {Goldstein}, J., \& {Zweibel}, E.~G. 2018, \mnras, 475,
  879

\bibitem[{{Townsend} \& {Teitler}(2013)}]{townsend2013}
{Townsend}, R.~H.~D. \& {Teitler}, S.~A. 2013, \mnras, 435, 3406

\bibitem[{{Triana} {et~al.}(2017){Triana}, {Corsaro}, {De Ridder}, {Bonanno},
  {P{\'e}rez Hern{\'a}ndez}, \& {Garc{\'{\i}}a}}]{triana2017}
{Triana}, S.~A., {Corsaro}, E., {De Ridder}, J., {et~al.} 2017, \aap, 602, A62

\end{thebibliography}
%

\begin{appendix}
\section{Rotational inversions}
\label{secrotinvapp}

    The angular dependence of each oscillation mode of a star can be expressed as a spherical harmonic, assuming spherical symmetry. Hence, modes are characterised by three numbers: $(n, \ell, m)$, which are the radial order, angular degree, and azimuthal order, respectively. Rotation leads to the breaking of degeneracy with respect to the azimuthal order, and as a consequence, instead of a single mode we can observe up to $2 \ell + 1$ modes with the same value $n$ and $\ell$ but different values of $m$. Only non-radial modes are subject to rotational splitting. The visibility of the different peaks due to the rotational splitting depends on the inclination of the rotation axis \citep{gizon2003}. To date, only dipole ($\ell=1$) modes have been used for rotational inversions for red giants. The challenges of using rotationally split $\ell=2$ modes have been  discussed by \cite{deheuvels2017}. So far, \cite{deheuvels2014} use $\ell=2$ modes only in the linear splitting approximation of \cite{goupil2013}. However, the information from dipole modes alone only allows determining the rotation rate of the core and the mean rotation rate of the convective envelope, and not at intermediate radii \citep{ahlborn2020}. Therefore, the focus of the present work is to study the determination of core and envelope rotation rates using dipole modes.
    
    To first order, the impact of rotation on the mode frequency $\omega_{nl}$ can be described as a linear perturbation to the mode frequency:
    \begin{align}
        \omega_{nlm}=\omega_{nl}+m\,\delta\omega_{nl},
    \end{align}
    with $\delta \omega_{nl}$ the rotational splitting \citep[e.g.][]{aerts2010}. For slow (when centrifugal forces can be ignored) shellular rotation, the splitting of a mode $(n, \ell)$ is given by:
    \begin{equation}
        \delta \omega_{nl} =\int_0^R \mathcal{K}_{nl}(r) \Omega(r) dr, \label{eqrot_split}
    \end{equation}
    where $r$ denotes the radial coordinate, $\Omega(r)$ is the radial rotation profile, and $\mathcal{K}_{nl}$ is an individual rotational kernel that is computed for each oscillation mode $(n,l)$ from the reference stellar model.

    Solving for $\Omega(r)$ in Eq.~(\ref{eqrot_split}) is known as a rotational inversion. Different methods that are used to solve for $\Omega(r)$ include RLS  methods \citep[e.g.,][]{gough1985, christensen1990}, MOLA inversions \citep{backus1968}, SOLA inversions \citep[SOLA][]{pijpers1992,pijpers1994} and most recently eMOLA inversions \citep{ahlborn2022}.
    
    The main idea of OLA methods is to linearly combine the individual rotational kernels $\mathcal{K}_i(r)$ using a set of inversion coefficients $c_i(r_0)$
    \begin{align}
        K(r,r_0)=\sum_{i\in\mathcal{M}}c_i(r_0)\,\mathcal{K}_i(r),\label{eqavgkernel}
    \end{align}
    to construct averaging kernels $K(r,r_0)$ localised at the target radii $r_0$ using the mode set $\mathcal{M}$. The inversion coefficients are chosen such that the averaging kernel is as localised as possible at the target radius. Given a localised averaging kernel the rotation at the target radius can be estimated by:
    \begin{align}
        \overline{\Omega}(r_0)=\int_{0}^{R}K(r,r_0)\,\Omega(r)\,\mathrm{d}r=\sum_{i\in\mathcal{M}}c_i(r_0)\,\delta\omega_i.
        \label{eqomegaavg}
    \end{align}
    The different OLA inversion methods (MOLA, SOLA, eMOLA) only differ in the way the inversion coefficients are determined. For all three methods, an objective function is minimised. In SOLA inversions, the coefficients are determined in order to minimise the squared difference between the averaging kernel and a target averaging kernel. For this work, we focus on eMOLA inversions only. For eMOLA \citep{ahlborn2022} the objective function reads:
    \begin{align}
        Z_\text{eMOLA}&=\int_0^RK(r,r_0)^2J(r,r_0)\,\text{d}r\nonumber\\
        &+\theta\left[\int_0^RK(r,r_0)J(r,r_0)\,\text{d}r\right]^2\nonumber\\
        &+\frac{\mu}{\mu_0}\underbrace{\sum_{i,j\in\mathcal{M}}c_i(r_0)c_j(r_0)E_{ij}}_{\sigma^2_{\Omega(r_0)}}.
        \label{eqobjeMOLA}
    \end{align}
    Here, $J(r,r_0)$ denotes a function that weights the amplitude of the averaging kernel depending on the distance to the target radius. A common choice is \citep[][]{backus1968,gough1985}
    \begin{align*}
        J(r,r_0)&=12(r-r_0)^2/R,
    \end{align*}
    where $R$ denotes the stellar radius. We denote the so-called trade-off parameter with $\mu$. This parameter balances the uncertainty of the solution from propagated errors $\sigma_{\Omega(r_0)}$ with the resolution of the inversions, as indicated by the width of the averaging kernels. By scaling the trade-off parameter with $\mu_0$ the third term in Eq.~(\ref{eqobjeMOLA}) becomes independent of the absolute value of the uncertainties, where $\mu_0$ given as,  
    \begin{align}
        \mu_0&=\frac{1}{M}\sum_{i\in\mathcal{M}}E_{ii},
    \end{align}
    with $\mathbf{E}$ denoting the variance-covariance matrix of the observed rotational splittings.

    The first term in Eq.~(\ref{eqobjeMOLA}) is also part of the original MOLA objective function. The second term, introduced in \cite{ahlborn2022}, penalises the accumulation of sensitivity in regions away from the target radius. To balance the original MOLA terms and the extension term, they introduce the second trade-off parameter $\theta$. For the calibration of the trade-off parameters $\mu$ and $\theta$ we refer to Sect.~3.3 and Appendix~B of \cite{ahlborn2022}.
\section{Grids of stellar models}\label{secstellarmodelsapp}

    \begin{table}
    \caption{Fundamental parameters of the two stellar model grids used.}             
    \label{tabgrids}
    \centering                          
    \begin{tabular}{lcccc}        
    \hline\hline                 
    Grid & Masses$/M_\odot$ & $Y_{\rm i}$ & $Z_{\rm i}$ & $\alpha_{\rm MLT}$ \\    
    \hline                        
       $M$ & [0.8, 2.0] & 0.2703 & 0.0142 & 1.8 \\ 
       $M, \alpha_{\rm MLT}$ & [0.8, 2.0] & 0.2703 & 0.0142 & [1.5, 2.0] \\
    \hline                                   
    \end{tabular}
    \tablefoot{Grid name, initial mass, initial helium mass fraction, initial metal mass fraction, and mixing length parameter for the models considered in this work. Square brackets indicate the range over which values are varied (see text for details).}
    \end{table}
    

    To search for reference models, we construct two different stellar model grids. We construct stellar evolutionary models using Modules for Experiments in Stellar Astrophysics \citep[MESA, version 12778,][]{paxton2011,paxton2013,paxton2015, paxton2018,paxton2019,jermyn2023}. To select the range of stellar parameters to consider in our study, we used the parameters of stars that have been analysed in terms of rotational inversions in \cite{deheuvels2012,dimauro2016} and \cite{triana2017}. The analysis of 16 red-giant stars presented in these studies suggests considering stellar masses of 0.8 up to 2~M$_\odot$. In our grids, we evolved the models up to a large frequency separations of 7~$\mu$Hz. We use the solar mixture of heavy elements by \cite{asplund2009}, the OPAL opacities from \cite{iglesias1996} extended by low temperature opacities from \cite{ferguson2005} and an Eddington-grey atmosphere \citep{eddington1926}. We use the {\tt mesa} equation of state. Convection is treated according to the mixing length theory \citep{boehm1958} with a free parameter $\alpha_{\rm MLT}$. We only vary mass ($M$) and the mixing-length parameter ($\alpha_{\rm MLT}$) to construct the different grids. We denote the grids as $M$-grid, where we only vary mass; and the $M, \alpha_{\rm MLT}$-grid, where we vary mass and the mixing length parameter. The details of the grids are summarised in Table~\ref{tabgrids}.
    
\begin{figure}
   \centering
   \includegraphics[width=\hsize]{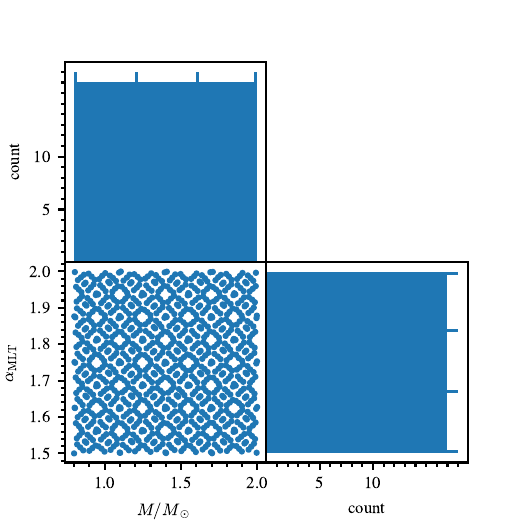}
      \caption{Parameter sampling of the mixing-length parameter $\alpha_{\rm MLT}$ and the stellar mass $M$ for the $M,\alpha_{\rm MLT}$-grid. In total, 1024 combinations of $M,\alpha_{\rm MLT}$ have been computed using a Sobol sequence. {\it Upper left:} Histogram of initial stellar masses. {\it Lower right:} Histogram of mixing length parameters. {\it Lower left:} Scatter plot of mixing-length parameter against mass.}
         \label{figsobol}
   \end{figure}

    To construct the $M$-grid, we have chosen the protosolar composition $Y=0.2704$ and $Z=0.0142$ of \cite{asplund2009} and a mixing length parameter of $\alpha_{\rm MLT}=1.8$. The mass is varied between 0.8 and 2.0~M$_\odot$ with a step of 0.01~M$_\odot$. For the $M, \alpha_{\rm MLT}$-grid, two parameters need to be varied simultaneously. To ensure a dense enough sampling of this two-dimensional parameter space, we used a two-dimensional Sobol sequence with a length of 1024 elements \citep{sobol1967} to draw parameter pairs $M, \alpha_{\rm MLT}$. This has been for example applied in \cite{bellinger2016} to sample the stellar model parameter space for their machine learning approach. The mixing length parameter $\alpha_{\rm MLT}$ is varied in a range between 1.5 and 2. An illustration of this process for the $M, \alpha_{\rm MLT}$-grid is shown in Fig.~\ref{figsobol}. The histograms show the near uniform sampling of both parameter axes. The scatter plot shows the quasi random distribution of the parameter pairs $(M, \alpha_{\rm MLT})$ avoiding holes or accumulations in certain regions of the parameter space. The composition is fixed to the solar values as for the $M$-grid.
\section{Synthetic data}\label{secsynthapp}
   
\subsection{Construction of synthetic mode sets}
    \label{secmodesets}
    We compute the oscillation frequencies and rotational kernels for given stellar models using the oscillation code GYRE \citep{townsend2013, townsend2018}. To mimic the set of observed frequencies of red-giant stars in the literature, we reduce the number of modes obtained from GYRE. For our synthetic mode sets, we only consider radial and dipole modes. Our synthetic mode sets consist of seven consecutive radial modes ($\ell=0$) centred around $\nu_{\rm max}$, the frequency of maximum oscillation power, and four consecutive radial orders of dipole modes ($\ell=1$) centred around $\nu_{\rm max}$ \citep[e.g.][App. A.3]{ahlborn2020}. We select the three dipole modes with the lowest mode inertia for each acoustic radial order. These mode sets are used in addition to the global observables as synthetic observables to search for the reference model for the rotational inversions. The construction of a realistic uncertainty model is discussed in Appendix~\ref{secuncert}.


\subsection{Synthetic rotational splittings}\label{secsynthrot}
To compute the rotational inversions, we also need a set of synthetic rotational splittings. We computed these rotational splittings for the dipole modes in our synthetic mode sets from Eq.~(\ref{eqrot_split}) given the rotational kernels from the selected stellar model as well as an imposed synthetic rotation profile. In this work, we used three different synthetic rotation profiles. We used two step-like profiles, featuring a step at the base of the convective envelope $r_{\rm rcb}$ and at 1.5 times the radius of the hydrogen burning shell $r_{\rm H}$. We refer to these models as the `envelope step' and  `core step' rotation profile. In addition, we used a synthetic rotation profile with a constant rotation below the base of the convective envelope and a power law decrease in the convective envelope. We refer to this profile as the `convective power law' profile \citep[see][for details of the synthetic rotation profiles]{klion2017}. The different synthetic rotation profiles are shown in Fig.~\ref{figrotprofiles}. We used $\Omega_{\rm core}/(2\pi)=750$~nHz and $\Omega_{\rm env}/(2\pi)=100$~nHz as default values. The impact of different synthetic rotation profiles on the rotational inversion result is demonstrated in \ref{secrotcomp}.
\subsection{Uncertainty model}
\label{secuncert}
\begin{figure}
   \centering
   \includegraphics[width=\hsize]{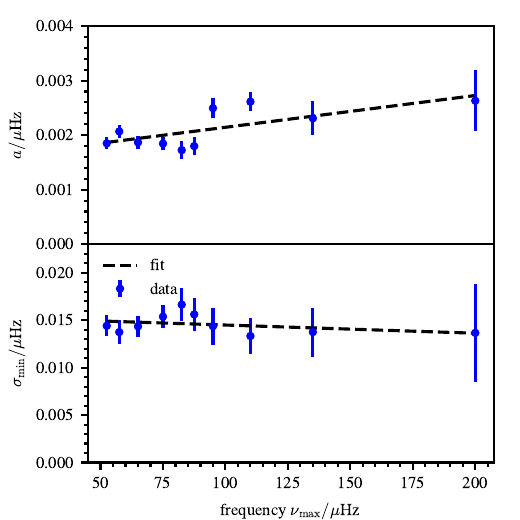}
      \caption{Values obtained for the parameters $a$ and $\sigma_{\rm min}$ as a function of $\nu_{\rm max}$ shown with blue points in the upper and lower panel, respectively. The error bars are obtained from the least squares fit to the uncertainties in the individual $\nu_{\rm max}$ bins. A linear fit to the data is shown with a black dashed line.}
         \label{figuncert_model}
   \end{figure}

    To mimic observed mode frequencies and rotational splittings, we need a realistic uncertainty model. These uncertainties are crucial for the best-fit model selection when computing realistic $\chi^2$ values as well as uncertainties on estimated rotation rates. Following \cite{schunker2016a} we describe the frequency uncertainties of an individual star as a function of mode frequency with a quadratic function with a minimum at $\nu_{\rm max}$:
    \begin{equation}
    \sigma_{\ell=0}(\nu) = \sigma_{\rm min}(\nu_{\rm max}) + a(\nu_{\rm max}) \cdot \left(\frac{\nu - \nu_{\rm max}}{\Delta\nu}\right)^2 ,
        \label{equncert_model}
    \end{equation}
    where $\sigma_{\rm min}(\nu_{\rm max})$ and $a(\nu_{\rm max})$ are parameters of the uncertainty model depending on the frequency of maximum oscillation power of the individual star. By dividing the frequency difference $\nu-\nu_{\rm max}$ by $\Delta\nu$, we introduce a typical scale for this difference. This removes the dependence of the parameters $a$ and $\sigma_{\rm min}$ on this typical frequency scale. We obtain the parameters $a$ and $\sigma_{\rm min}$ using 2237 red-giant stars analysed with the peak-bagging code TACO (Hekker et al. in prep., N. Themessl private communication). To determine the dependence of the uncertainty model parameters on the estimated $\nu_{\rm max}$, we group the stars in bins of $\nu_{\rm max}$ and fit our uncertainty model Eq.~(\ref{equncert_model}) to all radial mode frequency uncertainties in the $\nu_{\rm max}$ bin under consideration. The resulting parameter values of $a$ and $\sigma_{\rm min}$ as a function of $\nu_{\rm max}$ are shown in the upper and lower panel of Fig.~\ref{figuncert_model}, respectively. We model the parameter values as a linear function of $\nu_{\rm max}$, indicated by the black dashed lines in Fig.~\ref{figuncert_model}. We observe a slight increase of the parameter $a$ as $\nu_{\rm max}$  increases, while the minimum uncertainty stays approximately constant.
   
    For the sake of simplicity, we assume that the uncertainties on dipole mode frequencies are comparable to those of the radial modes. Therefore, we use the same model to compute the uncertainties on the dipole mode frequencies. Due to the smaller width of the dipole modes in the power spectrum as compared to radial modes, this assumption most likely overestimates the uncertainties on their frequencies, leading to an underestimation of the $\chi^2$ values. Further, the varying mixed mode nature has an impact on the uncertainties of the dipole mode frequencies. We discuss in Sect.~\ref{secuncertdiss} why the variation of the uncertainties with the mixed mode nature does not influence our results. Finally, from the assumption that uncertainties of modes with the same $(n, \ell)$ but different $m$ are equal, we derive the uncertainties of the rotational splittings from straight forward error propagation:
    \begin{equation}
        \sigma_{\delta \omega}(\nu)/(2\pi) = \frac{\sqrt{2}}{2} \cdot \sigma_{\ell=0}(\nu)
        \label{eqt_scale_split}
    \end{equation}
    This completes the construction of our synthetic data sets.
\section{Reference model selection}
\label{secbfmodelapp}
\subsection{Step 1: Selection based on global seismic properties}
\label{secglobselect}
    Due to the larger uncertainties of observed luminosities, surface gravities and effective temperatures, a pre-selection of the models based on these variables results in a too large number of stellar models to consider for the further computations. In contrast, the large frequency separation $\Delta\nu$ is known with high precision from the asteroseismic observations. As a second global asteroseismic quantity to put constraints on the reference model we use  the frequency of maximum oscillation power $\nu_{\rm max}$. For both the synthetic and reference stellar model, we use the large frequency separation and frequency of maximum oscillation power computed from scaling relations \citep{kjeldsen1995} with reference values from \cite{themessl2018}. This can be done in the same way for an observed star. The reference values in \cite{themessl2018} are calibrated to match the observed values of $\Delta\nu$ and $\nu_{\rm max}$ on the RGB. Therefore, as the first step of our selection procedure, we impose a maximum fractional error $f_x$ on the model value of the observable $x$ where $x=\Delta\nu, \nu_{\rm max}$:
   \begin{equation}
   \label{eqdeltanulim}
        \frac{|x_{\rm obs} - x_{\rm mod}|}{x_{\rm obs}} < f_{x}
   \end{equation}
   where we choose a value of $f_{\Delta\nu}=0.01$. For typical red-giant stars with asteroseismic observations, this results in a range of approximately $\pm0.1~\mu$Hz. For $\nu_{\rm max}$, we chose a threshold of $f_{\nu_{\rm max}}=0.1$, resulting in a range of approximately $\pm20~\mu$Hz. This is a rather generous limit to ensure that we do not exclude models \emph{a priori} from the selection.
   
   \subsection{Step 2: Selection based on radial modes}
   \label{secradselect}
   For the second step of the selection process, we use the radial mode $\chi_{\rm rad}^2$. For every model from the grid that passed the previous step, we compute radial modes with the oscillation code GYRE in the range $2.5 \cdot \sigma_{\rm env}$, where $\sigma_{\rm env}=0.28 \cdot \nu_{\rm max}^{0.88}$ \citep[see][]{mosser2012c}. Then we assign the radial mode frequencies of our synthetic observable set to frequencies of the potential reference model using the Hungarian algorithm \citep{kuhn1955}. The Hungarian algorithm solves the assignment problem between two sets, given a metric. In our case, we use the absolute frequency difference between observed and model frequencies as a metric. We note that the model frequencies need to be corrected for the surface effect when using observations of actual stars instead of synthetic observations. We use the radial mode $\chi_{\rm rad}^2$ as a measure of goodness of fit (see Eq.~\ref{eqchi2_definition}).
   Stellar models are then selected for the next step if the $\chi^2_{\rm rad}$ does not exceed a threshold, which is empirically selected. We have set this to a value of 500. Again, this is a conservative limit, and we do not consider a model with such a high value of $\chi^2_{\rm rad}$ to be a good fit to the observations. As the dipole mode properties are more important than radial modes for the rotational inversions, we do not want to exclude models based only on their radial mode properties. 

\subsection{The effect of the effective temperature}\label{secteffapp}
\begin{figure}
   \centering
   \includegraphics{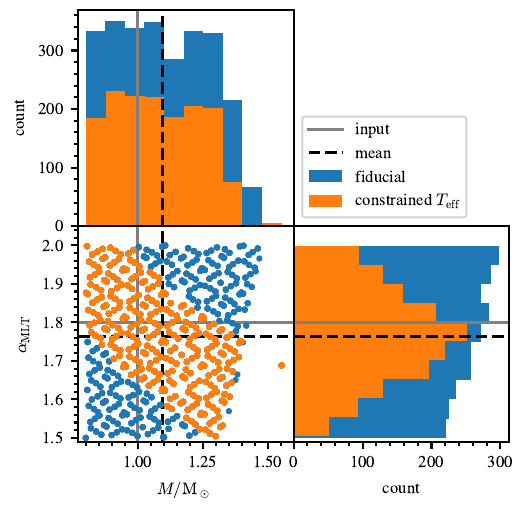}
      \caption{{\it Upper left:} Distribution of initial stellar masses for reference models selected based on global stellar properties and radial mode frequencies. The result for the fiducial model is shown in blue while the result including a constraint on the effective temperature is shown in orange. {\it Lower right:} Distribution of mixing length parameters for selected models. {\it Lower left:} Scatter plot showing the mixing length parameters versus the initial stellar mass for selected stellar models. The stellar models were selected following the selection process described in Appendix~\ref{secglobselect} and~\ref{secradselect} (blue points). Results including a limit on $T_{\rm eff}$ of $\pm100$~K around the observed value using the fiducial model are shown in orange. The models were selected from the $M, \alpha_{\rm MLT}$-grid. The grey lines and the black dashed lines refer to the input and the actual mean value of the blue distribution, respectively.}
         \label{figalphamassinterm}
   \end{figure}

For the results in Sect.~\ref{secensemble}, \ref{secsurfpertrot} and \ref{secparamdep} we have not used the effective temperature of the synthetic observations as a constraint on the set of reference models. In this subsection, we also use $T_{\rm eff}$ as a constraint on the reference models and discuss the impact on the rotation inversion results. In addition to $\Delta\nu$ and $\nu_{\rm max}$, we select models in a range of $\pm100$~K around the observed value in the first step of the selection procedure (see Appendix~\ref{secglobselect}). The resulting distributions of the initial stellar mass and the mixing length parameter $\alpha_{\rm MLT}$ are shown in the upper left and lower right panel of Fig.~\ref{figalphamassinterm} in orange. The results without constraints on the effective temperature as discussed in the main text are shown in blue. The mean value of the mass distribution when using the effective temperature as a constraint is $\overline{M}=1.08$ and the mean value of the mixing-length distribution is $\overline{\alpha}_{\rm MLT}=1.77$, not very different from the values obtained without adding the constraint. This shows that the two fundamental parameters that are varied on the grid are reproduced. The mean value of the mass is most likely biased towards higher values, as the grid is truncated at a mass of 0.8~M$_\odot$, which skews the distribution. The precision of the estimate could be further improved by choosing lower values of the threshold in $\chi^2_{\rm rad}$. We note however that it is not necessary to better constrain these fundamental parameters to obtain accurate rotation rates, as we discuss in Sect.~\ref{secparamdep}. The comparison of the orange and blue histograms of the mixing-length parameter show that this parameter is mostly constrained by the effective temperature. The restriction of the parameter space due to the different constraints also becomes apparent in the scatter plot shown in the lower left panel.
   
\subsection{Threshold values of $\rho$ and $\chi^2_{\rm dip}$}
\label{secthresh}
\begin{figure}
\centering
\includegraphics[]{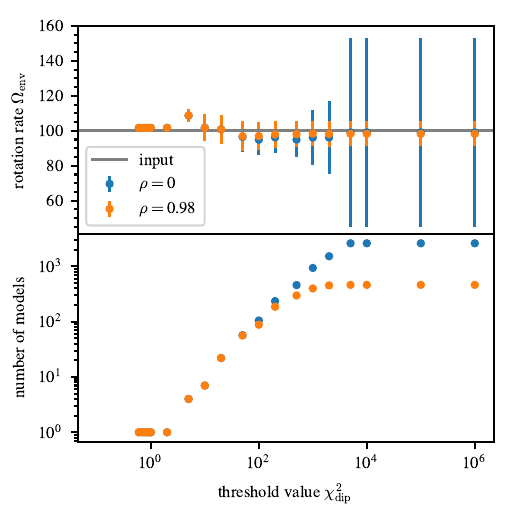}
\caption{{\it Upper panel:} Envelope rotation rate as a function of the threshold value in $\chi^2_{\rm dip}$ using the fiducial model as the synthetic observation. The reference models are selected from the $M,\alpha_{\rm MLT}$-grid. The input value is shown with the grey line. Threshold values of $\rho=0$ and $\rho=0.98$ are shown in blue and orange, respectively. {\it Lower panel:} Number of models selected as a function of the threshold value in $\chi^2_{\rm dip}$. Synthetic observation and stellar model grid are the same as in the upper panel. The colours have the same meaning as in the upper panel.}
\label{figthresh}
\end{figure}

For the ensemble inversion discussed in Sect.~\ref{secensemble} suitable threshold values for $\rho$ and $\chi^2_{\rm dip}$ need to be determined. These threshold values need to ensure that a large enough number of models is selected, and the rotation rate is estimated with small bias and reasonable uncertainties. This is shown in Fig.~\ref{figthresh}. In the upper panel, we show the estimated envelope rotation rate as a function of the threshold value in $\chi^2_{\rm dip}$ for two values of the threshold in $\rho$. The input value is indicated with the grey line. The fiducial model was used as the synthetic observation, and reference models were selected from the $M, \alpha_{\rm MLT}$-grid. The error bars indicate the reference model uncertainties as defined in Sect.~\ref{secensemble}. To exploit the ensemble inversion method, we aim at selecting a large enough number of reference models. This leads to larger values of the threshold in $\chi^2_{\rm dip}$ than one would use to optimise for a single reference model. In Fig.~\ref{figthresh}, the limiting cases for $\rho=0$ or large $\chi^2_{\rm dip}$ show that also using $\rho$ or $\chi^2_{\rm dip}$ alone would be sufficient to obtain an accurate estimate of the envelope rotation rate. The limiting cases are discussed in Appendix.~\ref{secmetricsapp} below.

In the lower panel of Fig.~\ref{figthresh} we show the number of models that got selected as a function of the threshold in $\chi^2_{\rm dip}$, for two values of the threshold in $\rho$. The number of models increases with increasing the threshold value of $\chi^2_{\rm dip}$ or decreasing values of $\rho$. For high values of $\chi^2_{\rm dip}$ the number of models saturates as all available models got selected. To obtain well constrained mean values, we argue that the standard error of the mean estimator goes as $\sigma/\sqrt{N}$, assuming normally distributed data with uniform standard deviations. So a number of selected models larger than about 50 leads to a robust estimator of the mean.

\subsection{Comparison of metrics}
\label{secmetricsapp}
\begin{figure*}
   \centering
   \includegraphics{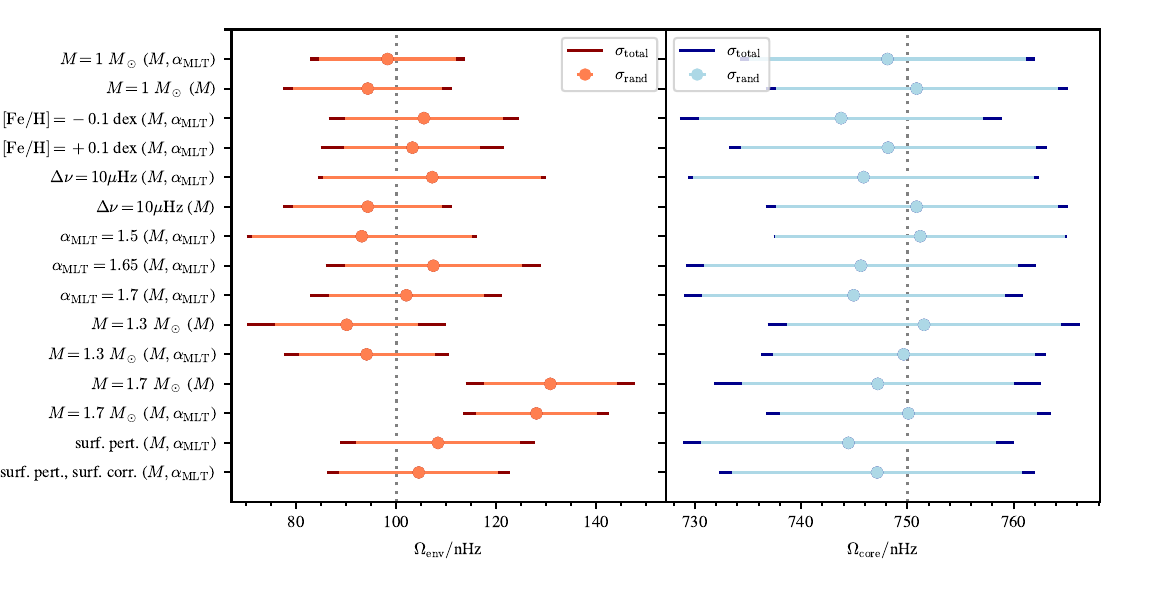}
      \caption{Ensemble inversion results for envelope (left) and core (right) rotation rates for different synthetic observations and using $\rho$ only as a metric. The numerical values are
summarised in Table~\ref{tabresultsrho}. The error given in the dark colours in each panel is calculated from the random and systematic errors in Table~\ref{tabresultsrho} by error propagation: $\sigma_{\rm total}=\sqrt{\sigma_{\rm random}^2+\sigma_{\rm ref}^2}$. The error bar in the light colours is representing the contribution of the random error $\sigma_{\rm rand}$ alone. The synthetic observation according to Table~\ref{tabsynobs} is given as the x-axis label. The vertical grey lines indicate the input values.}
         \label{figsummaryrho}
   \end{figure*}

\begin{figure*}
   \centering
   \includegraphics{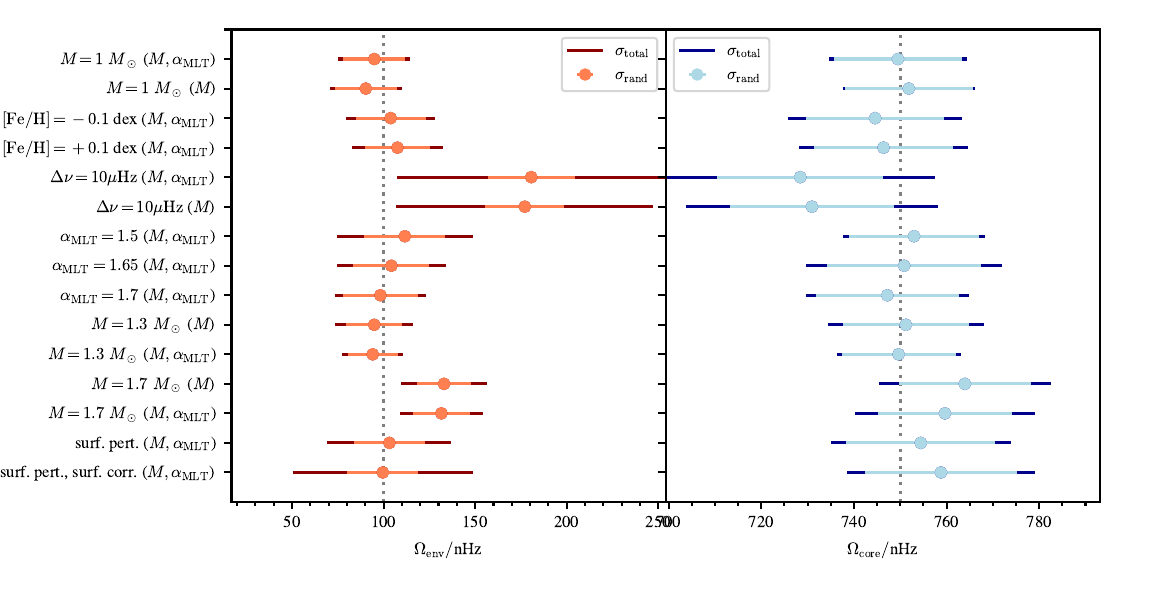}
      \caption{Ensemble inversion results for envelope (left) and core (right) rotation rates for different synthetic observations and using $\chi^2_{\rm dip}$ only as a metric. The numerical values are summarised in Table~\ref{tabresultschi2}. The error given in the dark colours in each panel is calculated from the random and systematic errors in Table~\ref{tabresultschi2} by error propagation: $\sigma_{\rm total}=\sqrt{\sigma_{\rm random}^2+\sigma_{\rm ref}^2}$. The error bar in the light colours is representing the contribution of the random error $\sigma_{\rm rand}$ alone. The synthetic observation according to Table~\ref{tabsynobs} is given as the x-axis label. The vertical grey lines indicate the input values.}
         \label{figsummarychi2}
   \end{figure*}

In this section we compare results computed with different metrics, that is with $\rho$ only (Table~\ref{tabresultsrho} and Fig.~\ref{figsummaryrho}) and with $\chi^2_{\rm dip}$  only (Table~\ref{tabresultschi2} and Fig.~\ref{figsummarychi2}). These should be compared to the results shown in Table~\ref{tabresults} and Fig.~\ref{figsummary}. When using $\rho$ only as a metric the rotation rates are estimated with a similar accuracy as in Fig.~\ref{figsummary}. However, the uncertainties increase slightly. The situation becomes worse when using the $\chi^2_{\rm dip}$ as the only metric. Here, the estimated rotation rates of the evolved model show larger deviations from the input values. In addition the uncertainties increase considerably. This makes it more difficult to differentiate different theories on the angular momentum transport. We hence conclude that both metrics should be applied simultaneously as described in the main text to obtain the most accurate and precise estimates of the internal rotation rates.

\begin{table*}
	\caption{Rotational inversion results for different synthetic observations using $\rho$ only.}
	\label{tabresultsrho}
	\centering
	\renewcommand{\arraystretch}{1.2}
	\begin{tabular}{c c c c c c c c c c}
		\hline\hline
		\rule{0pt}{12pt}name&$\Omega_{\rm core}$&$\sigma_{\rm rand}$&$\sigma_{\rm ref}$&$\Omega_{\rm env}$&$\sigma_{\rm rand}$&$\sigma_{\rm ref}$&$\rho_{\rm thresh}$&$\chi^2_{\rm dip, thresh}$&grid\\
		\hline
		\rule{0pt}{10pt}input&750&--&--&100&--&--&--&--&--\\
		\hline
		\rule{0pt}{10pt}$M=1~M_\odot$&748&13& 5&98&14& 8&0.98&--&$M, \alpha_{\rm MLT}$\\
		\hline
		\rule{0pt}{10pt}$M=1~M_\odot$&751&14& 6&94&15& 8&0.95&--&$M$\\
		\hline
		\rule{0pt}{10pt}$\left[{\rm Fe}/{\rm H}\right]=-0.1$ dex&744&14& 8&106&16&11&0.98&--&$M, \alpha_{\rm MLT}$\\
		\hline
		\rule{0pt}{10pt}$\left[{\rm Fe}/{\rm H}\right]=+0.1$ dex&748&14& 6&103&14&13&0.98&--&$M, \alpha_{\rm MLT}$\\
		\hline
		\rule{0pt}{10pt}$\Delta\nu=9~\mu{\rm Hz}$&746&17& 4&107&22& 7&0.98&--&$M, \alpha_{\rm MLT}$\\
		\hline
		\rule{0pt}{10pt}$\Delta\nu=9~\mu{\rm Hz}$&751&14& 6&94&15& 8&0.95&--&$M$\\
		\hline
		\rule{0pt}{10pt}$\alpha_{\rm MLT}=1.5$&751&14& 3&93&22& 7&0.98&--&$M, \alpha_{\rm MLT}$\\
		\hline
		\rule{0pt}{10pt}$\alpha_{\rm MLT}=1.65$&746&15& 8&107&18&13&0.98&--&$M, \alpha_{\rm MLT}$\\
		\hline
		\rule{0pt}{10pt}$\alpha_{\rm MLT}=1.7$&745&15& 8&102&16&12&0.98&--&$M, \alpha_{\rm MLT}$\\
		\hline
		\rule{0pt}{10pt}$M=1.3~M_\odot$&752&13& 8&90&15&14&0.95&--&$M$\\
		\hline
		\rule{0pt}{10pt}$M=1.3~M_\odot$&750&13& 6&94&14&10&0.98&--&$M, \alpha_{\rm MLT}$\\
		\hline
		\rule{0pt}{10pt}$M=1.7~M_\odot$&747&13& 9&131&14&11&0.95&--&$M$\\
		\hline
		\rule{0pt}{10pt}$M=1.7~M_\odot$&750&13& 6&128&13& 9&0.98&--&$M, \alpha_{\rm MLT}$\\
		\hline
		\rule{0pt}{10pt}surf. pert.&744&14& 8&108&17&11&0.98&--&$M, \alpha_{\rm MLT}$\\
		\hline
		\rule{0pt}{10pt}surf. pert., surf. corr.&747&14& 6&105&16&10&0.98&--&$M, \alpha_{\rm MLT}$\\
		\hline
	\end{tabular}
	\tablefoot{The rotational inversion results are computed with the ensemble rotational inversion described in Sect.~\ref{secensemble}. All rotation rates and uncertainties given in units of nHz.}
\end{table*}

\newpage

\begin{table*}
	\caption{Rotational inversion results for different synthetic observations using $\chi^2_{\rm dip}$ only.}
	\label{tabresultschi2}
	\centering
	\renewcommand{\arraystretch}{1.2}
	\begin{tabular}{c c c c c c c c c c}
		\hline\hline
		\rule{0pt}{12pt}name&$\Omega_{\rm core}$&$\sigma_{\rm rand}$&$\sigma_{\rm ref}$&$\Omega_{\rm env}$&$\sigma_{\rm rand}$&$\sigma_{\rm ref}$&$\rho_{\rm thresh}$&$\chi^2_{\rm dip, thresh}$&grid\\
		\hline
		\rule{0pt}{10pt}input&750&--&--&100&--&--&--&--&--\\
		\hline
		\rule{0pt}{10pt}$M=1~M_\odot$&750&14& 6&95&18&10&--&500&$M, \alpha_{\rm MLT}$\\
		\hline
		\rule{0pt}{10pt}$M=1~M_\odot$&752&14& 4&90&17&11&--&500&$M$\\
		\hline
		\rule{0pt}{10pt}$\left[{\rm Fe}/{\rm H}\right]=-0.1$ dex&745&15&12&104&20&15&--&500&$M, \alpha_{\rm MLT}$\\
		\hline
		\rule{0pt}{10pt}$\left[{\rm Fe}/{\rm H}\right]=+0.1$ dex&746&15&11&108&18&18&--&500&$M, \alpha_{\rm MLT}$\\
		\hline
		\rule{0pt}{10pt}$\Delta\nu=9~\mu{\rm Hz}$&728&18&23&181&24&70&--&100&$M, \alpha_{\rm MLT}$\\
		\hline
		\rule{0pt}{10pt}$\Delta\nu=9~\mu{\rm Hz}$&731&18&21&177&22&67&--&100&$M$\\
		\hline
		\rule{0pt}{10pt}$\alpha_{\rm MLT}=1.5$&753&15& 6&112&23&31&--&500&$M, \alpha_{\rm MLT}$\\
		\hline
		\rule{0pt}{10pt}$\alpha_{\rm MLT}=1.65$&751&17&14&104&21&22&--&500&$M, \alpha_{\rm MLT}$\\
		\hline
		\rule{0pt}{10pt}$\alpha_{\rm MLT}=1.7$&747&16& 9&98&21&15&--&500&$M, \alpha_{\rm MLT}$\\
		\hline
		\rule{0pt}{10pt}$M=1.3~M_\odot$&751&14&10&95&16&16&--&500&$M$\\
		\hline
		\rule{0pt}{10pt}$M=1.3~M_\odot$&750&13& 6&94&14&10&--&500&$M, \alpha_{\rm MLT}$\\
		\hline
		\rule{0pt}{10pt}$M=1.7~M_\odot$&764&15&12&133&15&19&--&1000&$M$\\
		\hline
		\rule{0pt}{10pt}$M=1.7~M_\odot$&760&15&13&132&16&17&--&1000&$M, \alpha_{\rm MLT}$\\
		\hline
		\rule{0pt}{10pt}surf. pert.&754&17&11&103&20&28&--&500&$M, \alpha_{\rm MLT}$\\
		\hline
		\rule{0pt}{10pt}surf. pert., surf. corr.&759&17&12&100&20&46&--&500&$M, \alpha_{\rm MLT}$\\
		\hline
	\end{tabular}
	\tablefoot{The rotational inversion results are computed with the ensemble rotational inversion described in Sect.~\ref{secensemble}. All rotation rates and uncertainties given in units of nHz.}
\end{table*}

\end{appendix}

\end{document}